\documentclass{aa}  

\usepackage{color}
\usepackage{graphicx}

\usepackage{mathtools}
\usepackage{siunitx}
\usepackage[bottom]{footmisc}
\usepackage{tablefootnote}
\usepackage[flushleft]{threeparttable}
\usepackage{adjustbox}
\usepackage{subcaption}
\usepackage{float}

\usepackage{txfonts}
\begin{document}

   \title{The phase-space distribution of the M\,81  satellite system}
   \author{Oliver Müller
          \inst{1}          
          \and
          Nick Heesters\inst{1}
                    \and 
          Marcel S. Pawlowski\inst{2}
                    \and 
          Kosuke Jamie Kanehisa\inst{2,3}
          \and 
          Federico Lelli\inst{4}
          \and 
          Noam I. Libeskind\inst{2}
          }

   \institute{Institute of Physics, Laboratory of Astrophysics, Ecole Polytechnique Fédérale de Lausanne (EPFL), 1290 Sauverny, Switzerland\\
              \email{oliver.muller@epfl.ch}
         \and
             Leibniz-Institut für Astrophysik Potsdam (AIP), An der Sternwarte 16, D-14482 Potsdam, Germany
             \and
             Institut für Physik und Astronomie, Universität Potsdam, Karl-Liebknecht-Straße 24/25, 14476 Potsdam, Germany
             \and
             INAF, Arcetri Astrophysical Observatory, Largo Enrico Fermi 5, I-50125, Florence, Italy
             }

   \date{Received tba; accepted tba}

% \abstract{}{}{}{}{} 
% 5 {} token are mandatory
 
  \abstract
   {The spatial distribution of dwarf galaxies around their host galaxies is a critical test for the standard model of cosmology because it probes the dynamics of dark matter halos and is independent of the internal baryonic processes of galaxies. Co-moving planes-of-satellites have been found around the Milky Way, the Andromeda galaxy, and the nearby Cen\,A galaxy, which seem to be at odds with the standard model of galaxy formation. Another nearby galaxy group, with a putative flattened distribution of dwarf galaxies, is the M\,81 group. In this paper, we present a quantitative analysis of the distribution of the M\,81 satellites using a Hough transform to detect linear structures. Using this method, we confirm a flattened distribution of dwarf galaxies. Depending on the morphological type, we find a minor-to-major axis ratio of the satellite distribution to be 0.5 (all types) or 0.3 (dSph), which is in line with previous results for the M\,81 group. Comparing the orientation of this flattened structure in 3D with the surrounding large-scale matter distribution, we find a strong alignment with the local sheet and the planes-of-satellites around the Andromeda galaxy and Cen\,A.  Furthermore, the satellite system seems to be lopsided. Employing line-of-sight velocities for a subsample of the dwarfs, we find no signal of co-rotation. Comparing the flattening and motion of the M\,81 dwarf galaxy system with TNG50 of the IllustrisTNG suite we find good agreement between observations and simulations, {but caution that i) velocity information of half of the satellite population is still missing, ii) current velocities are coming mainly from dwarf irregulars clustered around NGC 3077, which may hint towards an infall of a dwarf galaxy group and iii) some of the dwarfs in our sample may actually be tidal dwarf galaxies. From the missing velocities, we predict that the observed frequency within IllustrisTNG may still range between 2 to 29 per cent.  Any final conclusions about the agreement/disagreement with cosmological models needs to wait for a more complete picture of the dwarf galaxy system.} %Because the M\,81 group is a dynamically complex system, we do not draw 
   
   }

   \keywords{Galaxies: dwarf; galaxies: groups: individual: M81; galaxies: distances and redshifts; cosmology: large-scale structure of Universe.
               }

   \maketitle
%
%________________________________________________________________

\section{Introduction}
The distribution and motion of dwarf galaxies around the Milky Way \citep{1976MNRAS.174..695L,2008ApJ...680..287M,2012MNRAS.423.1109P,taibi2023portrait} and the Andromeda galaxy \citep{2006AJ....131.1405K,2006MNRAS.365..902M,2013Natur.493...62I} has sparked an on-going debate on whether they pose a challenge to the current $\Lambda$+Cold Dark Matter ($\Lambda$CDM) standard model of cosmology (see e.g. \citealt{2005A&A...431..517K,2005MNRAS.363..146L,
2007MNRAS.374...16L,2012MNRAS.423.1109P,2014ApJ...784L...6I,2015MNRAS.452.3838C,Pawlowski2020,2022arXiv220502860S} and the reviews by \citealt{Pawlowski2018,2021NatAs...5.1185P}). 
The structure and kinematics of satellite systems pose a strong test for $\Lambda$CDM because they are driven by the gravitational dynamics on scales of hundreds of kpc, so they do not depend on internal baryonic processes \citep{Pawlowski2018,2018Sci...359..534M}.
This has pushed several teams to seek for similar structures outside the Local Group  \citep{2014Natur.511..563I,MuellerTRGB2018,2021ApJ...917L..18P,2021A&A...654A.161H}. Clear evidence was found for our neighbor Cen\,A and its satellite system, showing a statistically significant correlation in phase-space \citep{2015ApJ...802L..25T,Muller2016,2018Sci...359..534M,Muller2021b,2023MNRAS.519.6184K}. For the Sculptor group, \citet{2021A&A...652A..48M} pointed out a flattened distribution of satellites, as well as some signs of coherent motion, but the system needs additional follow-up observations of their velocities to assess the situation. For other systems in the Local Volume -- a sphere of 10\,Mpc around our point of view -- claims of flattened distributions have been made \citep{2017A&A...602A.119M,MuellerTRGB2018}. Intriguingly, most flattened structures seem to be aligned with the local cosmic web \citep{2015MNRAS.452.1052L,2019MNRAS.490.3786L}. This may point towards a common formation scenario for these structures.

A nearby galaxy group 
%in the Local Volume 
with a well studied satellite population is the M\,81 group of galaxies at a distance of $\approx$3.7\,Mpc \citep{2000ApJS..128..431F,2013AJ....146..126C}. The central region of the group is characterized by three major galaxies (M81, M82 and NGC 3077) that are interacting with each other, as demonstrated by a complex HI network of filaments, tidal tails { and candidate tidal dwarf galaxies } \citep{1994Natur.372..530Y,2008AJ....135.1983C,2008ApJ...689..160W}. 
Due to dynamical friction, it has been argued that such compact arrangements of three galaxies are unlikely be found in a $\Lambda$CDM cosmology \citep{2017MNRAS.467..273O}. 

The M\,81 group has been a target for dwarf galaxy searches by different teams. \citet{2009AJ....137.3009C} used the Canada France Hawaii Telescope (CFHT) telescope to survey a large field of 65 deg$^2$, discovering 22 dwarf galaxy candidates, of which 14 were confirmed based on Hubble Space Telescope (HST) follow-up observations \citep{2013AJ....146..126C}. More recently, the group has been studied with the Hyper Suprime Cam (HSC). \citet{2015ApJ...809L...1O,2019ApJ...884..128O} surveyed a field of 6.5 deg$^2$, discovering two additional faint dwarf galaxies. Finally, \citet{2022ApJ...937L...3B} combined archival HSC data (seven $\approx$1.5deg$^2$ fields) and found six dwarf galaxy candidates. 

In the seminal work by \citet{2013AJ....146..126C}, the authors noted that the satellite structure seemed to be flattened, with an on-sky $rms$ thickness of 61\,kpc at the distance of M\,81, when considering the dwarf spheroidals of the satellites, which is also well-aligned with the local sheet. In this work, we aim to provide a quantitative analysis of the spatial and kinematic properties of the M\,81 group and a first {-- but not final --} comparison to $\Lambda$CDM simulations of this system. In Sec.\,\ref{data}, we discuss the data and methods we are using, in Sec.\,\ref{distribution} we provide a discussion on the distribution and motion of the satellite system, in Sec.\,\ref{simulations} we compare the observations to cosmological simulations {and discuss major caveats}, and in Sec.\,\ref{summary} we provide a summary and conclusions.

\section{Data and Methods}
\label{data}

\begin{figure}[ht]
    \centering
    \includegraphics[width=\linewidth]{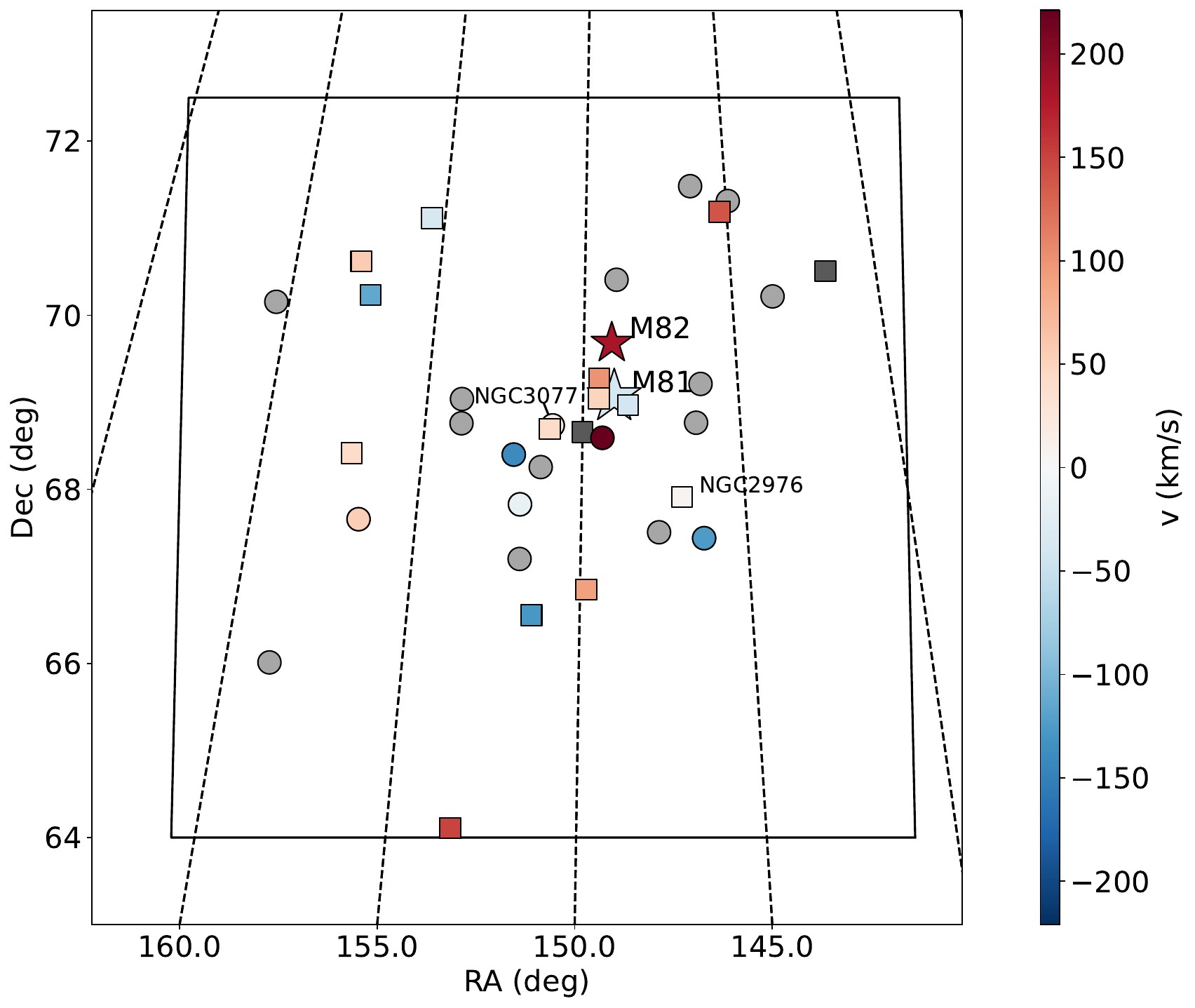}
    \caption{The survey footprint from \citet{2009AJ....137.3009C,2013AJ....146..126C}. The two main galaxies M\,81 and M\,82 are indicated as stars, the dwarf spheroidals as dots, the dwarf irregulars as squares. The two brightest satellites (NGC\,2976 and NGC\,3077) are marked. Colors indicate the observed velocities, gray stands for no velocity measurement currently available.
    }
    \label{fig:field}
\end{figure}

\begin{figure*}[ht]
    \centering

        \includegraphics[width=0.49\linewidth]{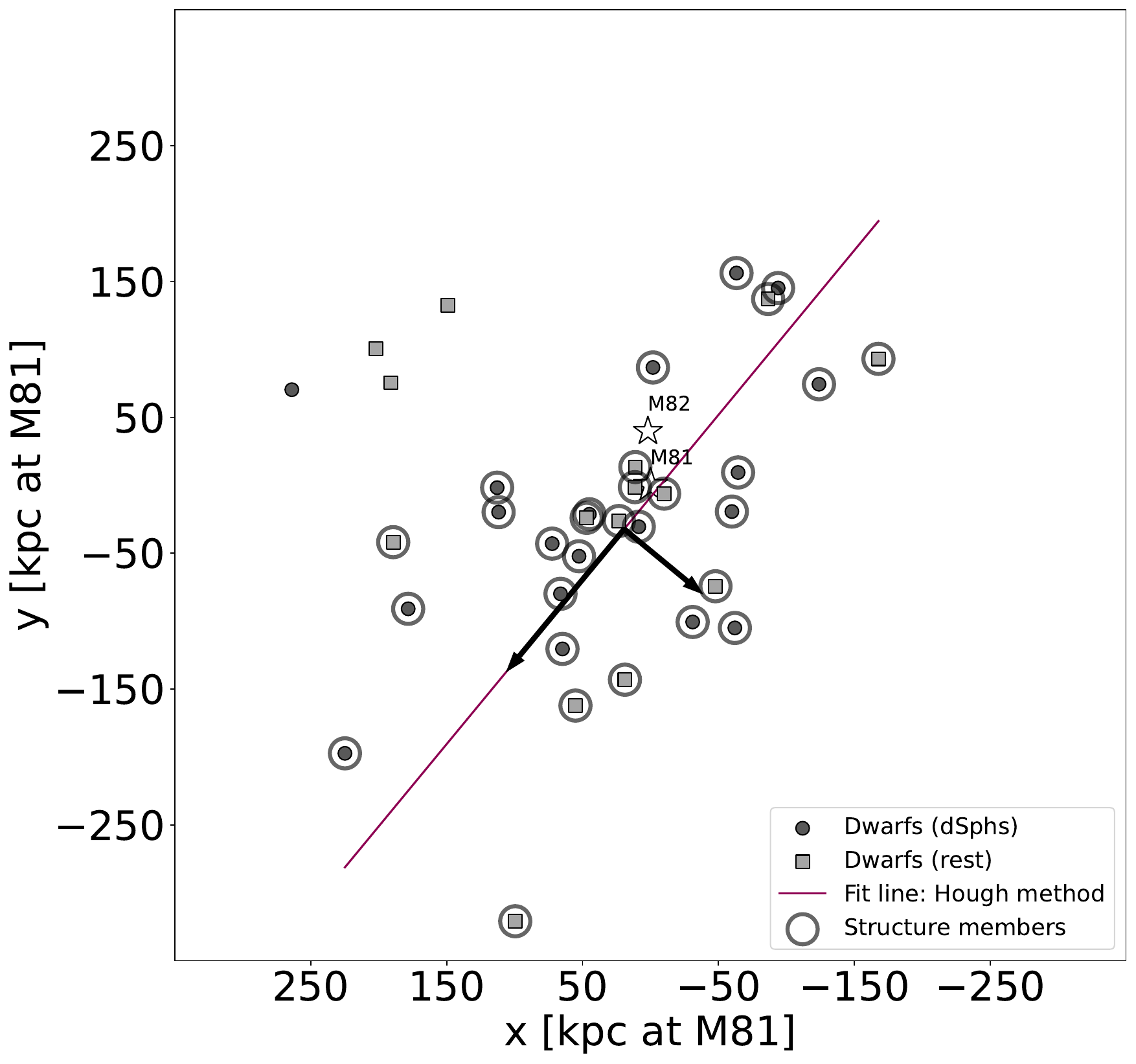}
        \includegraphics[width=0.31\linewidth]{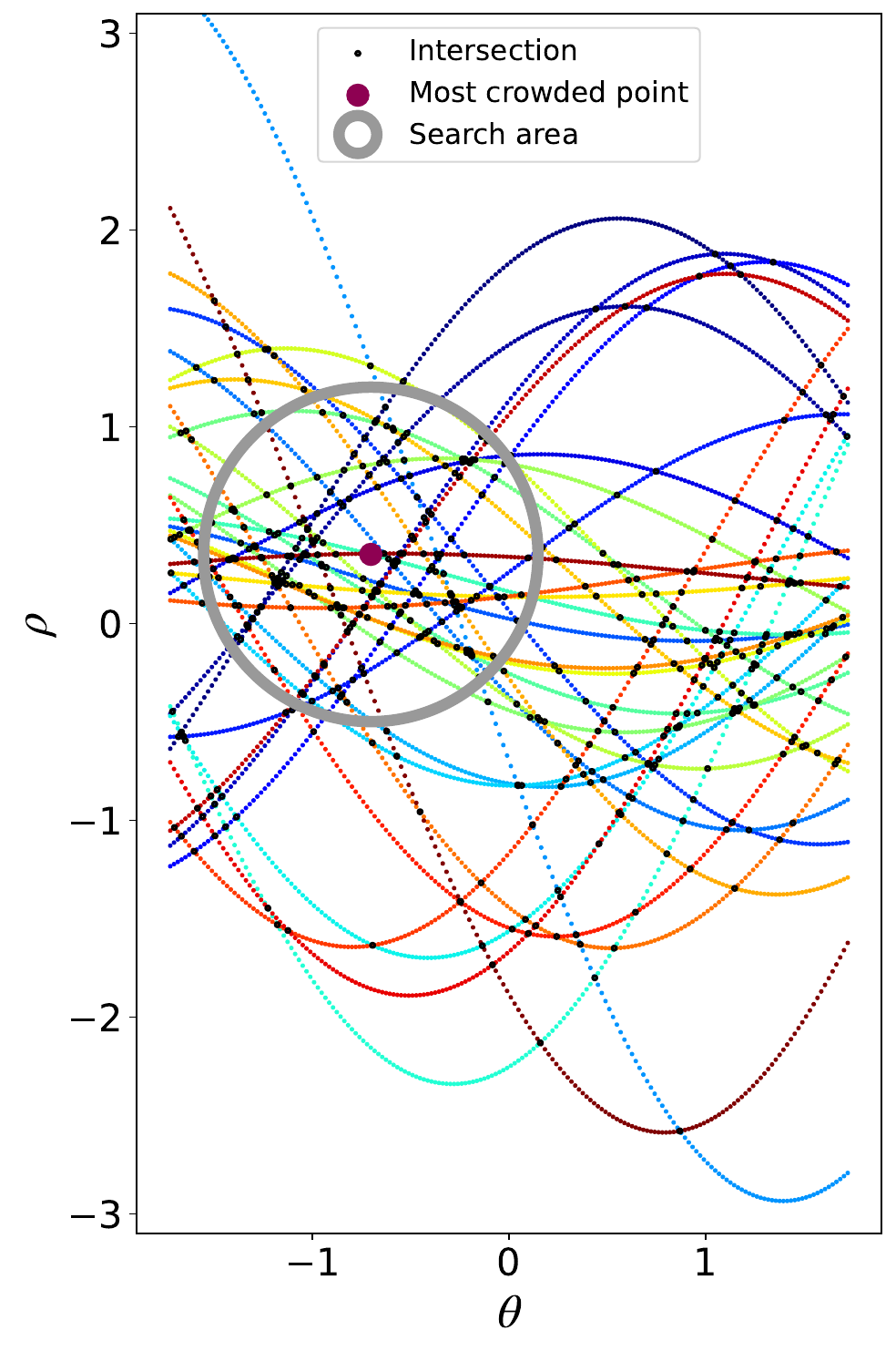}\\
        \includegraphics[width=0.49\linewidth]{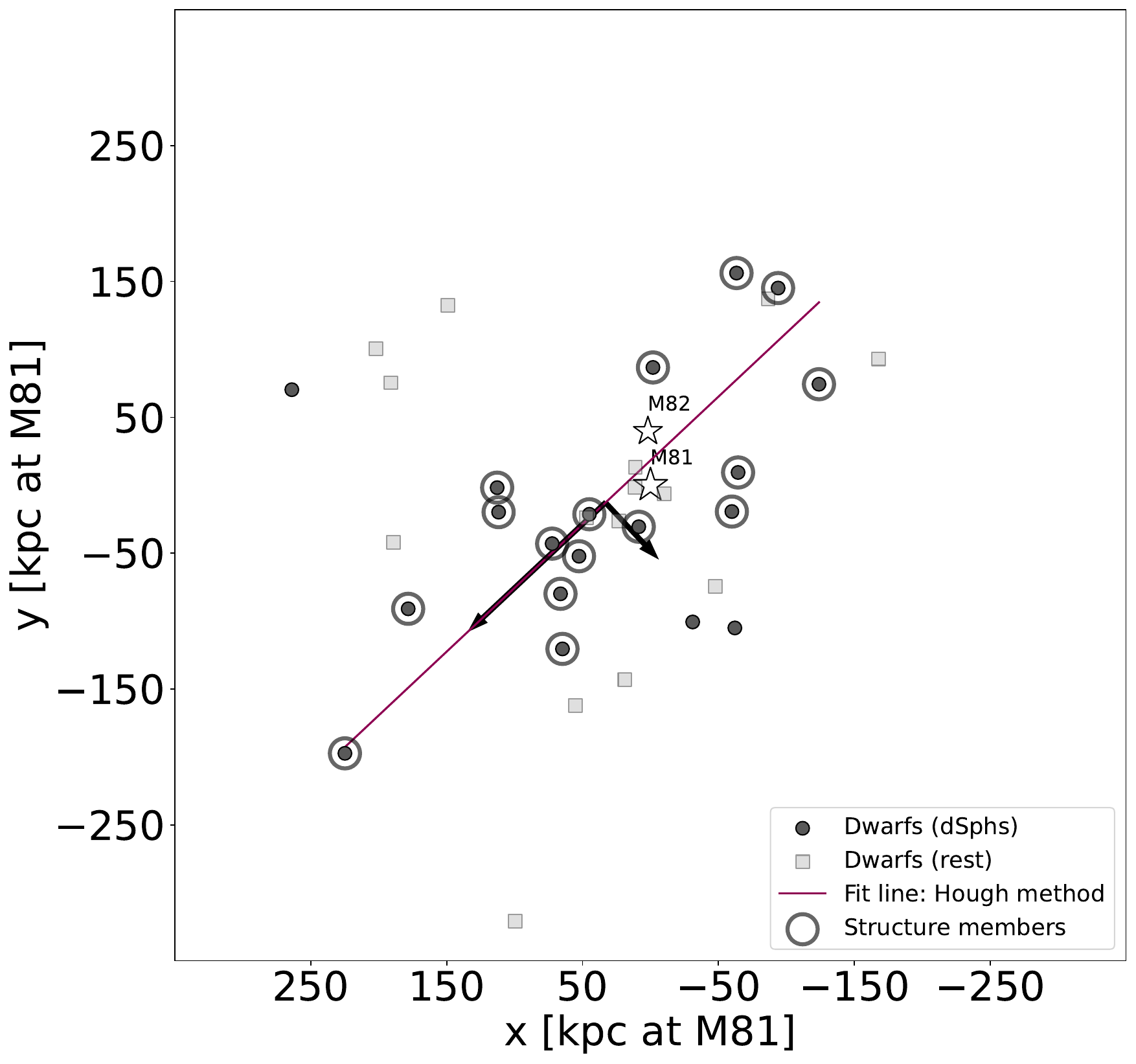}
   \includegraphics[width=0.31\linewidth]{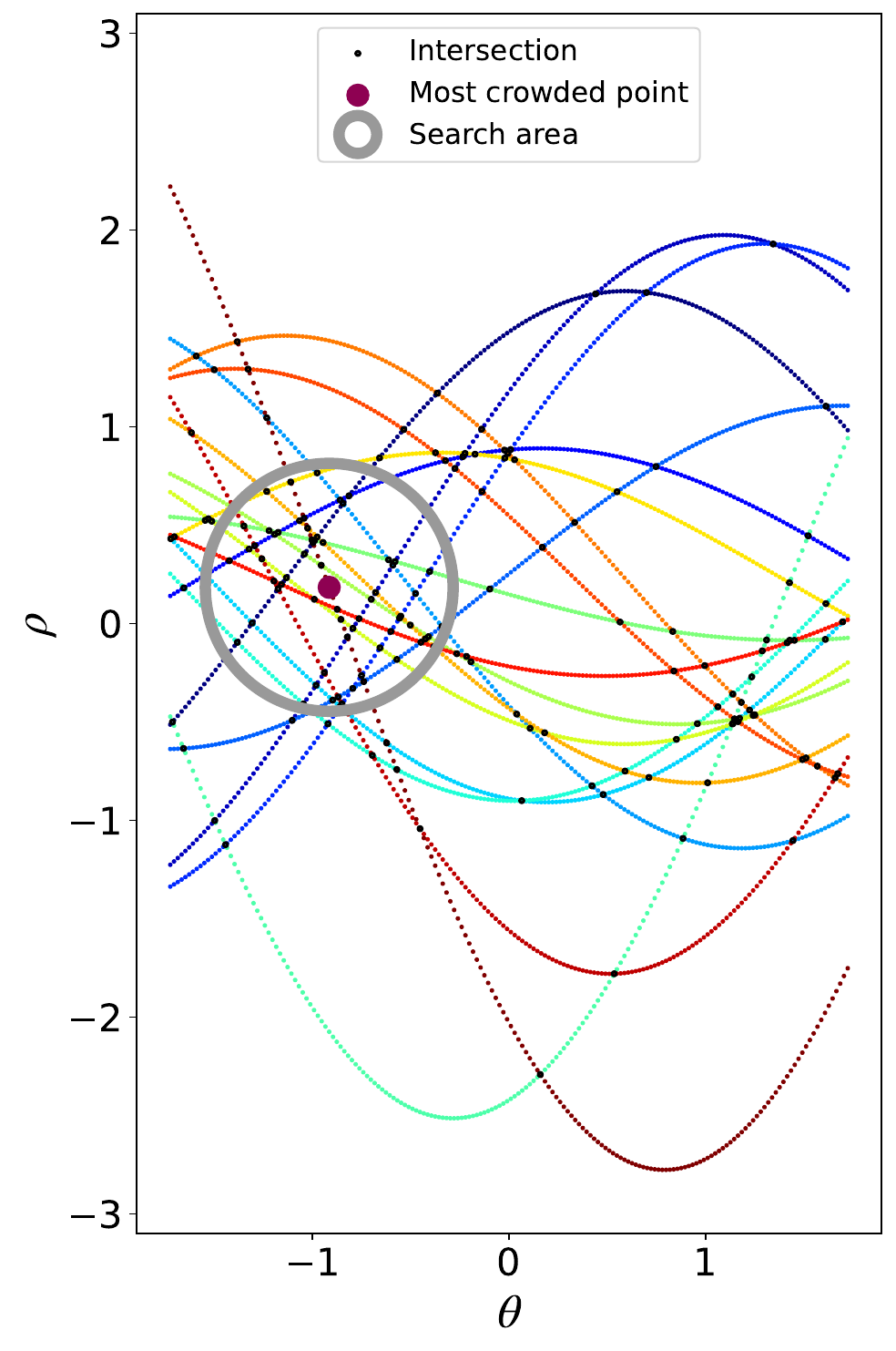}
%    \includegraphics[width=0.49\linewidth]{M81_Hough_all.pdf}
%        \includegraphics[width=0.49\linewidth]{M81_Hough_dSphonly.pdf}\\
%   \includegraphics[width=0.4\linewidth]{M81_hough_space.pdf}
%   \hspace{20pt} %\hspace{40pt}
%   \includegraphics[width=0.4\linewidth]{M81_hough_space_dSph_only.pdf}
    \caption{Left: The M81 group satellite system in relative coordinates, with M81 (star) at the origin and indicating projected separations at the distance of M\,81 in kpc. M\,82 is shown as an empty star. The dwarf satellites of different morphologies are shown as circles and squares. We emphasize the abundance of dwarf spheroidals (dSphs) as black circles compared to other morphologies as gray squares. The best-fit line is determined via the Hough method, which finds all but four satellites (upper left corner) to be members (circled gray) of a flattened structure. The $rms$ dimensions of this structure are indicated with black arrows. Top: the results using all morphological types of satellites, Bottom: only using the dSphs.
    Right: The M\,81 satellite system in the Hough parameter space. The parameter $\rho$ plotted on the y-axis is the normal distance from the fit line to the origin, while $\theta$ is the angle between this $\rho$ vector and the x-axis in the image space. Each curve corresponds to a series of possible parameter pairs for one dwarf. The curves are z-score normalized along each axis, i.e., they have a mean of zero and a standard deviation of one. The red dot indicates the parameter pair at the center of the densest crossing region (grey circle), which is optimized to maximize the number of structure members and the flatness of the system simultaneously. The curves which do not cross the grey region stem from the declared outliers in the scatter plot left.
    }
    \label{fig:M81}
\end{figure*}

We use the catalog of group members from \citet{2013AJ....146..126C}, which includes their own HST follow-up observations of the dwarf galaxy candidates found in a CFHT survey of 65 deg$^2$ \citep{2009AJ....137.3009C}, as well as literature data. We consider galaxies with $<-18.0$ in absolute magnitudes in the $r$-band as satellites. This effectively removes {the two gravitationally dominant galaxies} M\,81 and M\,82 from the list, but still includes the bright galaxies NGC\,2976 (with $M_r=-18.0$\,mag) and NGC\,3077 (with $M_r=-17.8$\,mag). M\,81 and M\,82 have stellar luminosities (derived from the $K_s$ band) of $10^{10.95}$\,M$_{\odot}$ and  $10^{10.59}$\,M$_{\odot}$ \citep{2013AJ....145..101K}, respectively.  To date, the \citet{2013AJ....146..126C} data represents the most complete survey of the M\,81 group and its surroundings, going beyond the second turnaround radius of M\,81 (which is $\approx$230\,kpc). Because \citet{2013AJ....146..126C} do not compile the distances of the literature data, we have taken the values from the online version of the Local Volume catalog \citep{2013AJ....145..101K}\footnote{https://relay.sao.ru/lv/lvgdb/, last accessed 15.09.2022.}. A recent deep survey by \citet{2022ApJ...937L...3B} uncovered another six ultra-faint dwarf galaxies, which are clustered around NGC\,3077. We refrain from adding these to our list of satellites, because their HSC survey covered only 50\,kpc around M\,81, which would bias any study of the distribution of dwarf galaxies in this group. In Table\,\ref{tab:sample} we provide the list of dwarf galaxies used in our study.

How may the survey and its footprint bias the study of the dwarf galaxy population? The M\,81 group lies in a region in the sky where galactic cirrus is dominant and this may cause trouble with assessing the distribution of the satellites. However, \citet{2013AJ....146..126C} made extensive tests with artificial star detection to probe their selection criteria and found that their detection of dwarf galaxies is not biased by cirrus (see their Fig.\,33 for the cirrus overlaid on top of their survey field). It is evident that, even in regions with strong cirrus, dwarf galaxy candidates were detected. Another issue may arise from the survey footprint. The flattened distribution found by \citet{2013AJ....146..126C} is aligned with the diagonal of the footprint, see Fig.\,\ref{fig:field}. Because the diagonals will maximize the radial distance for which dwarfs can be found, it might prefer finding linear structures along these lines. To mitigate that, we would need to extend the survey footprint and see where the dwarf galaxy detection drops to the background. However, there seems to be a drop of dwarf galaxy detections towards the border of the survey footprint, which indicates that the dwarf galaxy population is sampled well enough towards the edges. We note though that some known galaxies like NGC\,2403 or UGC4483 are generally considered to be members of the M\,81 group \citep{2005astro.ph..9207K}, but are well outside of the survey footprint, with projected separations of $\sim$15 deg (corresponding to roughly 1\,Mpc). Here, we do not consider them as satellites of the M\,81 group.

\begin{figure*}[ht]
    %\centering
    \includegraphics[width=0.49\linewidth]{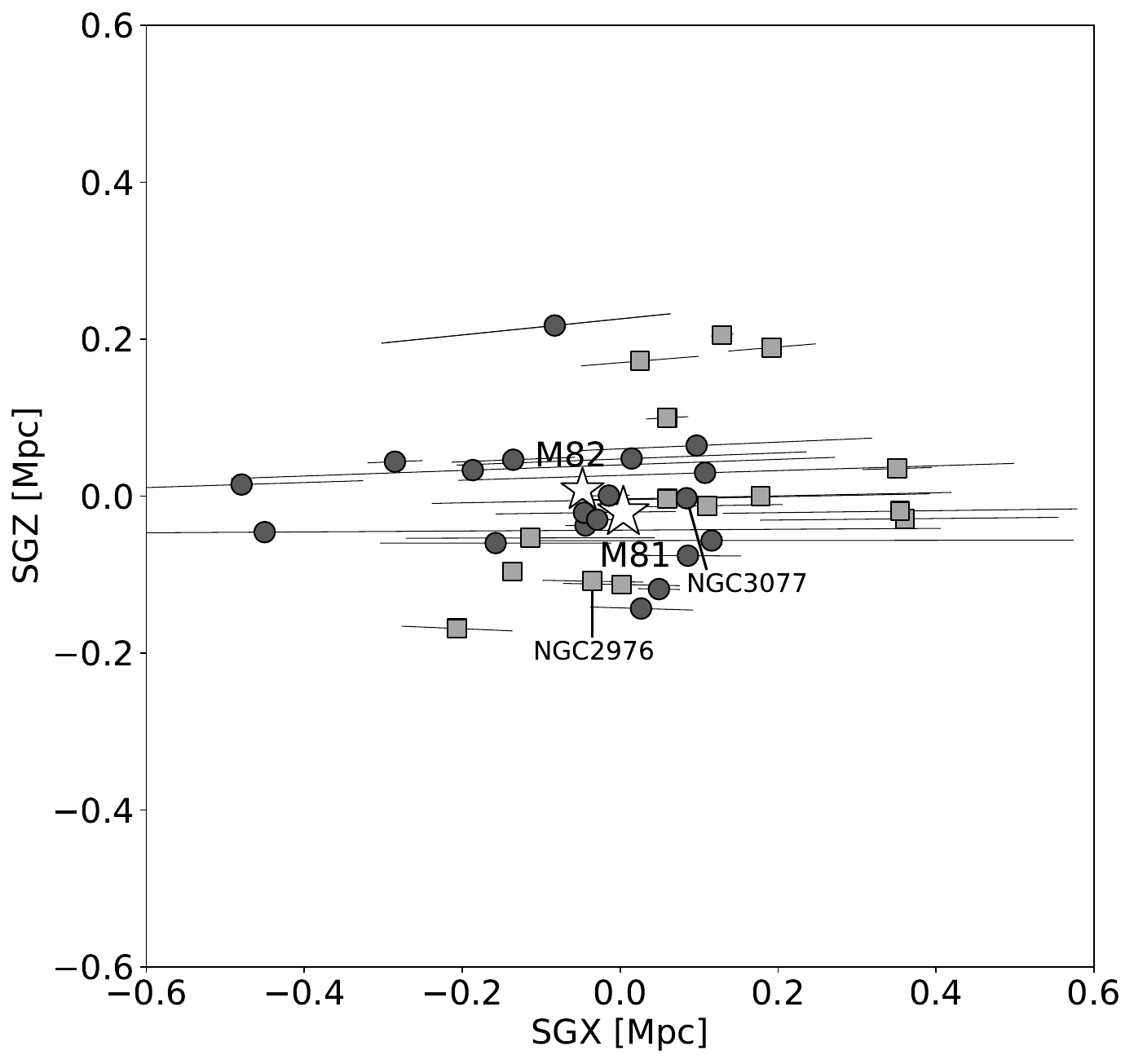}
    \includegraphics[width=0.49\linewidth]{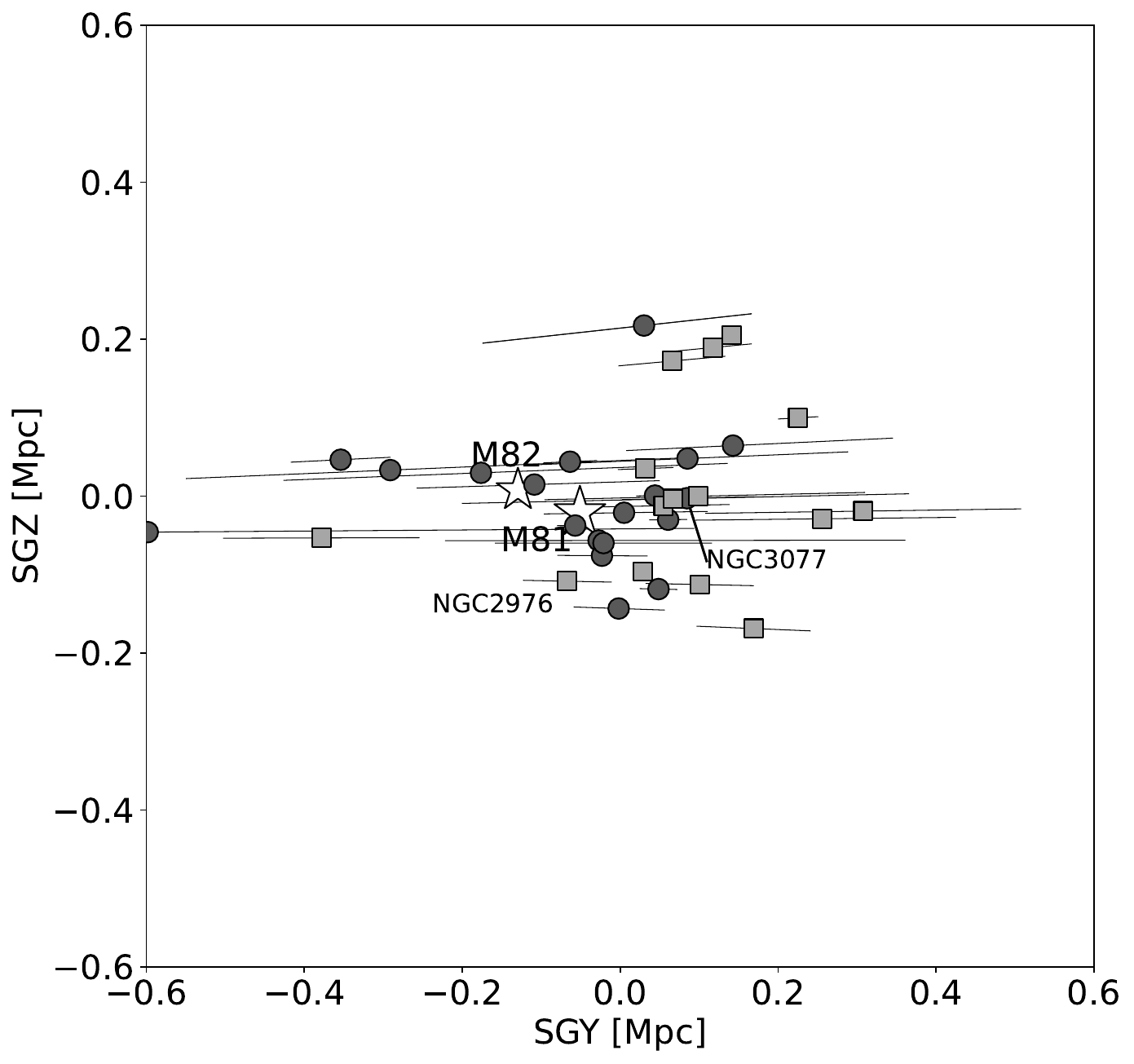}\\
    \includegraphics[width=0.49\linewidth]{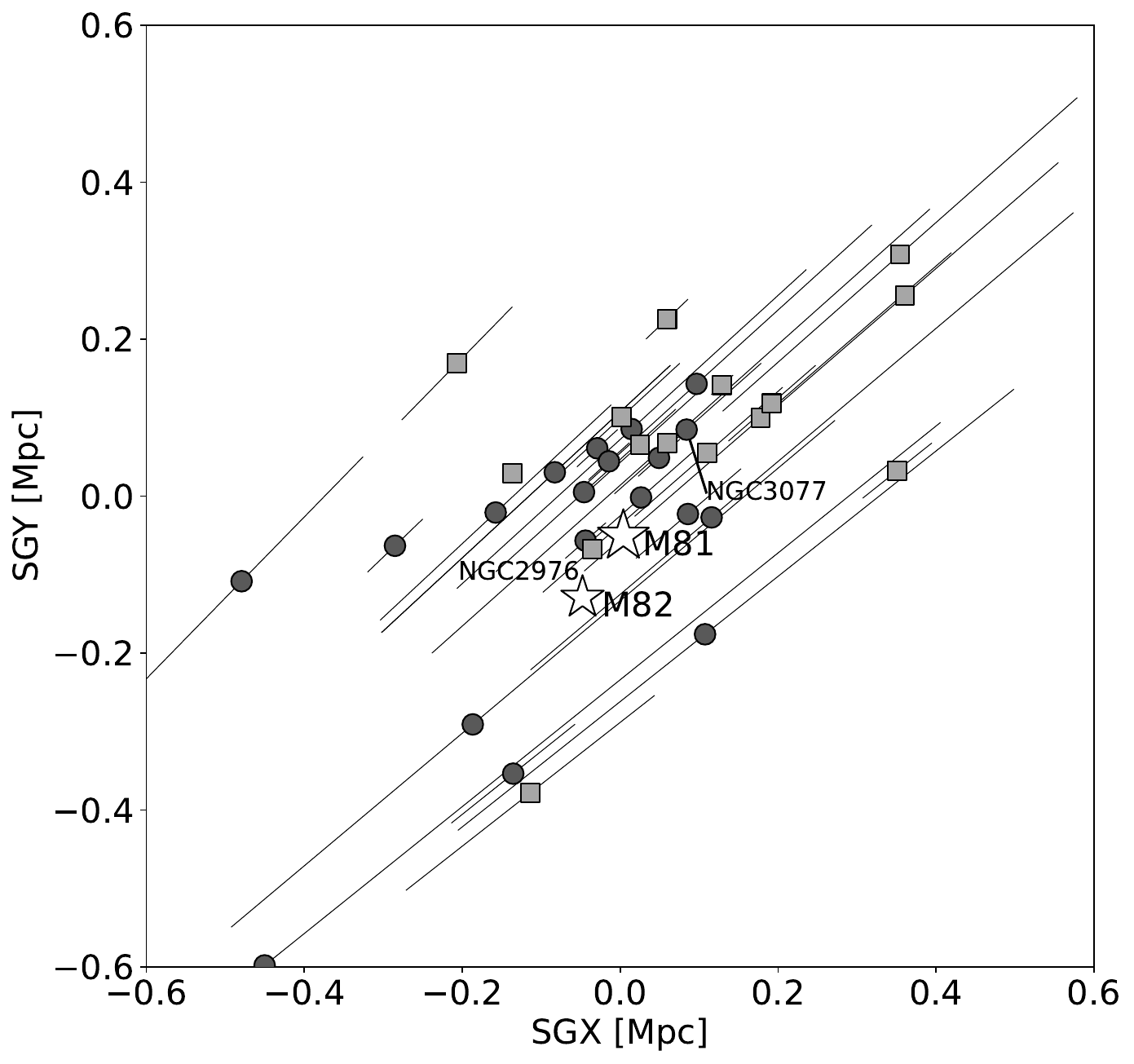}
    \includegraphics[width=0.49\linewidth]{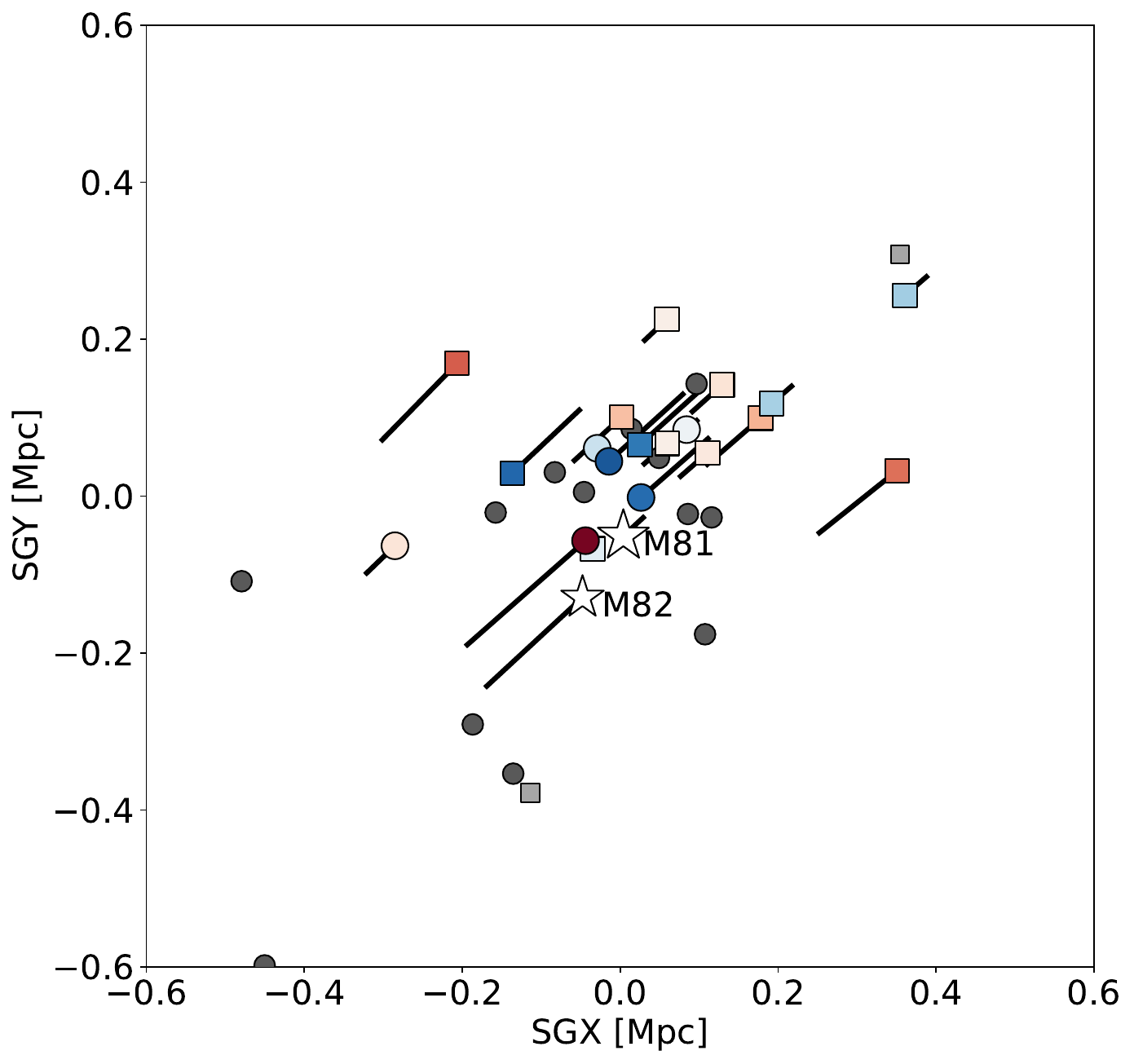}
    \caption{The satellite system of M\,81 in Cartesian Supergalactic coordinates, shifted to the center of the satellite distribution. The dSphs are indicated in dark grey dots, the dIrr as squares in a lighter grey, and the main galaxies M\,81 and M\,82 as white stars. The thin black lines indicate the distance uncertainties. The top panels represent an edge-on view of the plane, the bottom panels a face-on view. In the bottom right panel the velocities are indicated as thick lines, as well as the color, with red indicating galaxies moving away from us in respect to the mean velocity of the system, and blue towards us. The color-coding is the same as in Fig\,\ref{fig:pv}.}
    \label{fig:SG}
\end{figure*}

To study the flattening of the satellite distribution, we employed the Hough transformation as described in \citet{2021A&A...654A.161H}. In short, the Hough transform \citep{hough1959machine,hough1962method} was originally developed as a tool to detect straight lines and other simple shapes in digital images. The principle of the method is a voting system in which all points in an image decide on the best-fit parameters from a predefined set of slope and intercept pairs. {For every data point we define a range of lines that cross the point from different directions. With these lines the point votes for the best fit that should describe the overall linear distribution. Since the image space has no bounds for slope and intercept, the lines are transformed to the so-called Hough space via the relation 
\begin{equation}
    \rho = x\,\cos(\theta) + y\,sin(\theta).
\label{hough_formula}
\end{equation}
Here, x and y are the coordinates of the data point in the image space. The parameter $\rho$ is the orthogonal distance from the origin to the line passing through the data point. Finally, $\theta$ is the angle between the $\rho$ vector and the x-axis. In this space, the parameter ranges are closed with $\theta \in [-\pi/2,\pi/2]$ and $\rho \in [-d, d]$, where $d$ is the diagonal of the image. The votes by each data point are generated by calculating the values for $\rho$ corresponding to a range of $\theta$ values via equation \ref{hough_formula}. This process amounts to a dotted curve in the Hough space for every data point. Each of these dots represents one vote this is, one line that passes through one of the data points.} The region {in the Hough space} where the highest number of such lines cross represents the parameters with the highest number of votes at the end of this process, optimally describing the data set at hand. In an ideal scenario, where all data points are arranged on a perfectly straight line, there is a single crossing point in this parameter space. For scattered data points, however, the curves no longer cross in a single point. For this case we identify the region with the highest over-density of crossing lines and adopt a variable search area, allowing us to probe different scales. We optimize the area such that the structure flatness and the number of voting members are maximized simultaneously. 
This method allows for an educated estimation of the members of a potentially correlated satellite structure, i.e. a plane close to seen edge-on, in the absence of three dimensional information. The success of the method has been demonstrated on the M31 system \citep{2021A&A...654A.161H}, for which only about half of the satellites appear to be part of a phase-space correlation \citep{2013Natur.493...62I}. Details about this fitting technique can be found in \citet{2021A&A...654A.161H}.

\section{Satellite distribution}
\label{distribution}

{In this section we look into the spatial and kinematic distribution of the satellite system of M\,81.}

\subsection{Planes of satellites}

In Figure\,\ref{fig:M81} we present the {2D} distribution of the galaxies in the M\,81 group and the morphological types of the satellites. {A simple total least square (TLS) fit of the satellite system {in 2D} reveals an axis ratio of $b/a\,=\,0.79$ (semi minor axis $b\,=\,91$\,kpc, semi major axis $a\,=\,115$\,kpc), but this approach does not consider the possibility that there may be outliers not belonging to a flattened structure, so artificially increasing the $b/a$ ratio.} The Hough method identifies a significantly flattened structure along the diagonal of the field. Out of the 34 satellites, 30 follow this elongated structure. Of the four satellites (HS117, d1028+70, DDO82, d1042+70) not being part, only one is a dwarf spheroidal (d1042+70). Removing these four outliers, we measure a minor-to-major axis ratio $b/a=0.50$ (semi minor axis $b=61$\,kpc, semi major axis $a=122$\,kpc), which is similar to what is found for Cen\,A ($b/a\approx$0.5, \citealt{Muller2016}), but spatially less flattened than the Local Group planes (we note that the Local Group planes are studied in 3D due to better distance and thus 3D position accuracy). It is also consistent with the $rms$ width estimated by \citet{2013AJ....146..126C}. If we repeat the steps (i.e. Hough transformation and removing the outliers) for the dSph only, we get $b/a\,=\,0.34$ (semi minor axis $b\,=\,42$\,kpc, semi major axis $a\,=\,124$\,kpc), with 16 out of 19 dSph belonging to the flattened structure. However, two dwarf galaxies -- KK77 and F8D1 -- which were previously considered in the Hough fitting, are now not found to be part of the flattened structure. This is interesting, because F8D1 has a disrupted profile exhibiting a tidal tail \citep{2023MNRAS.518.2497Z}  which is aligned approximately along the minor axis of the flattened structure. If we consider this shape as a tracer of the motion of the dwarf, it is moving out from the planar structure.
{If we consider the entire dSph sample in a TLS fit, we measure an axis ratio of $b/a\,=\,0.62$ (semi minor axis $b\,=\,73$\,kpc, semi major axis $a\,=\,117$\,kpc).} That the flattening is higher for the dSph population compared to the total population is in line with the finding of \citet{2013AJ....146..126C}, who suggested there exists a flattened structure consisting of the dwarf spheroidals in the M\,81 group.

\begin{figure}[ht]
    \centering
    \includegraphics[width=\linewidth]{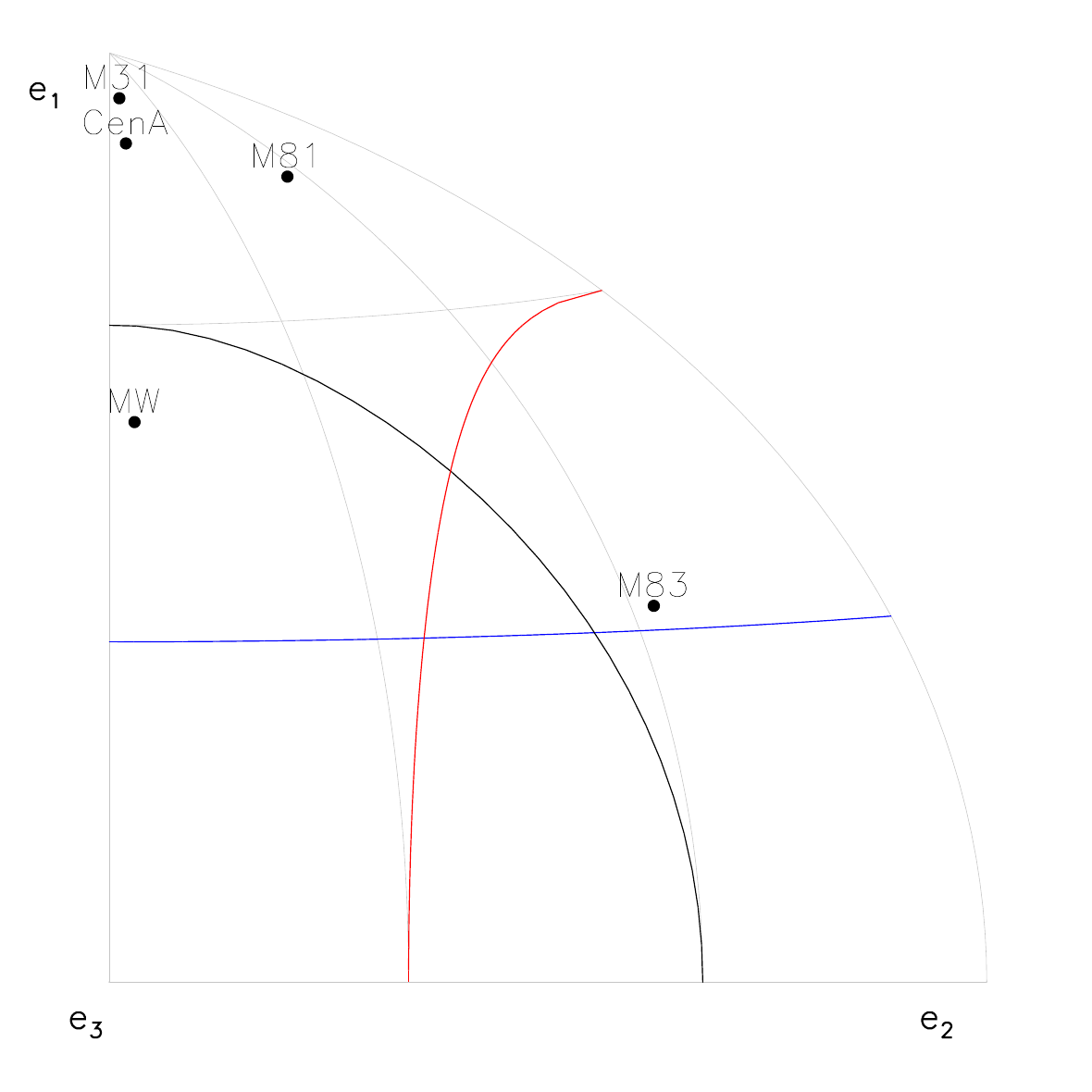}
    \caption{The orientation of the poles of planes-of-satellites with respect to the surrounding large-scale matter distribution. The eigenvector $\hat{e}_{1}$ corresponds to the direction of fastest collapse and is pointing towards the local void. The three lines (black, red, and blue) correspond a cone of 60 degrees off from the eigenvectors.}
    \label{fig:orientation}
\end{figure}

How is the flattened structure oriented in respect to the local universe? One way of answering this question is to examine how the  flattened structure is aligned with respect to the surrounding large-scale matter distribution, as characterized by the velocity-shear tensor, specifically its principle eigenvectors \citep[e.g][]{2012MNRAS.425.2049H}. The eigenvector corresponding to slowest collapse, ($\hat{e}_{3}$), defines the spines of cosmic filaments whereas the axis of greatest compression ($\hat{e}_{1}$) is associated with the normal of cosmic sheets \citep[see][and references therein]{2018MNRAS.473.1195L}. \citet{2015MNRAS.452.1052L,2019MNRAS.490.3786L} performed such an analysis on the known dwarf galaxy planes (i.e. MW, M31, Centaurus A) {in 3D} using a quasi-linear reconstruction of the Local Universe from CF2 \citep{2018NatAs...2..680H,2013AJ....146...86T} and found that many of their normals tend to be  closely aligned with $\hat{e}_{1}$, the eigenvectors of the shear tensor corresponding to the axis of greatest compression.  To test whether the M\,81 satellite population is aligned with any of these vectors, we {can} transform the system into supergalactic coordinates {\citep{1956VA......2.1584D,1998astro.ph..9343L}. In the supergalactic coordinates the Local Sheet -- including the Local Group, NGC253, Cen\,A, and M\,81 -- is aligned with the supergalactic SGX-SGY plane, which corresponds to the  $\hat{e}_{2}$-$\hat{e}_{3}$ plane.} %{We assume that the planar structure is seen edge-on, this is, the 3D normal is aligned with the plane of the sky. From the Hough fit of the full sample (dSphs and dIrr) we get the normal in equatorial coordinates ([$+$0.7712, $+$0.6366, $+$0.0000]).}  %The necessary steps are explained in \citet{Muller2016} and consist of applying rotation matrices to the normal of the satellite distribution. 
The satellite system in {in 3D} supergalactic coordinates is presented in  Fig.\,\ref{fig:SG}.
{We studied the alignment of the satellite system with a principal component analysis (pca) using the python package \textit{sklearn} \citep{scikit-learn}. The pca is a method to find the eigenvectors of a data set. Here, we use it to estimate the direction with the smallest variance, which for a flattened distribution of data points corresponds to its normal vector.
Considering all dwarfs -- including the four outliers which we deem not be part of the plane-of-satellites -- the normal for the planar structure  in supergalactic coordinates is [$-$0.1314, $+$0.0374, $+$0.9906].  The normal  including all dwarf galaxies in the flattened structure is [$-$0.0899, $+$0.0772, $+$0.9930]. The angle between the two normals is only 3 degrees, so they align well and can be regarded as equal. The position of the normal in respect to the shear tensor is given in Fig.\,\ref{fig:orientation}. 

The flattened satellite structure of M\,81 has its normal aligned close to $\hat{e}_1$ of the velocity-shear tensor defined by the large-scale matter distribution in the local universe. This is consistent with \citet{2013AJ....146..126C} who noted ``an apparent flattening of the distribution of gas-deficient systems to the supergalactic equatorial plane". However, we note that employing the Hough transformation we find both an alignment considering dSphs only, and dSphs and dIrrs together. 
The alignment of M\,81's satellite system with the Local Sheet is also for the satellite systems of M\,31 and Cen\,A \citep{2015MNRAS.452.1052L,2019MNRAS.490.3786L}. However, for the best studied plane-of-satellite system, this is, around the Milky Way no such alignment can be found. This may indicate that these structures have different origins.
}

It is interesting to note that the four dwarf galaxies (HS117, d1028+70, DDO82, d1042+70) not being part of the main structure form some kind of elongated spur as well, which is going parallel to the main structure. It has a minor-to-major axis ratio $b/a=0.28$ (semi minor axis $b=13$\,kpc, semi major axis $a=46$\,kpc). There is a gap of 222\,kpc between the best-fitting plane from these four dwarfs and the main flattened stucture. This gap is well visible in 2D in Fig.\,\ref{fig:field} and in 3D in Fig.\,\ref{fig:SG}. 

\subsection{Lopsidedness}
{Do we find evidence for an asymmetric distribution of the satellite system? This can be studied using different metrics testing for lopsidedness in the group. Here, we use two simple approaches, these are, a) the distance of the centroid from M\,81, and b) a hemisphere approach.

{In the first approach, we test whether the center of the satellite distribution coincided with the dominant galaxy M\,81.
We define the center or centroid of the Hough fit as 

\begin{equation}
\mathbf{r}_{0} = \frac{1}{N} \sum_{i=1}^{N} \mathbf{r}_{i}.
\label{centroid}
\end{equation}

Here $N$ is the number of structure members and $\mathbf{r}_{i}$ are the member coordinates.}
The centers for the two Hough fit lines (using all dwarfs and using only the dSph) are presented in Fig\,\ref{fig:M81}. They are separated from each other by 23\,kpc. The centroids for both samples are off from the position of M\,81 by $\approx$40\,kpc. Is this significant? To test this, we run a Monte Carlo simulation where we generated for each iteration 30 random positions uniformely drawn within the box defined by the plane-of-satellites (i.e. with $b=61$\,kpc and $a=122$\,kpc). Finding the center of the point cloud for each iteration, we estimate that an offset of 40\,kpc happens in only 0.2\% of the cases, in other words, is significant at the 3$\sigma$ level.  This may indicate that the dwarf satellite population is not symmetrically distributed around M\,81, which could be a physical effect caused by the on-going interaction between M81, M82 and NGC3077. It could, also arise from an infalling group of dwarfs, see the  subsection \ref{sub:kinematics} for such hints. 

Another way to test lopsidedness is splitting the distribution of the satellites in two hemipsheres. Such lopsidedness has been found within the Local Group \citep{2013ApJ...766..120C}, as well as statistically in galaxy groups in the SDSS survey \citep{2016ApJ...830..121L}. A study by \citet{2017ApJ...850..132P}  of these SDSS systems in $\Lambda$CDM cosmological simulations has found that this anisotropy is not in tension with observations. Kanehisa et al. (submitted) however finds the M31 satellite system's lopsidedness to be significantly in tension with $\Lambda$CDM simulation expectations. Therefore it is interesting to see whether we find this in the M\,81 group or not.
Taking the minor axis of the satellite distribution as a separation line, which we fix at M\,81's position (the brightest galaxy in the group), we find that 11 satellite galaxies are on one side and 23 on the other. Assuming a Bernouilli distribution for a satellite being on one side or another (i.e. a fair coin flip), we estimate that finding 11 or less satellites on one side has a probability of 5.8\% (note that we have to consider two cases, i.e. finding 11 or less and finding 23 or more in a Bernouilli experiment). This is not passing the 3$\sigma$ threshold and we therefore do not find it to be significant.

%One could assume that a lopsidedness arises from the overlap of the satellite populations of M\,81 and M\,82. However, the lopsidedness found here is on the other side of M\,81 compared to M\,82. This is clearly visible in Fig\,\ref{fig:SG}, bottom left panel. It could, however, arise from an infalling group of dwarfs, see the next subsection for such hints. 

}
\subsection{Kinematics}
\label{sub:kinematics}

\begin{figure}[ht]
    \centering
    \includegraphics[width=\linewidth]{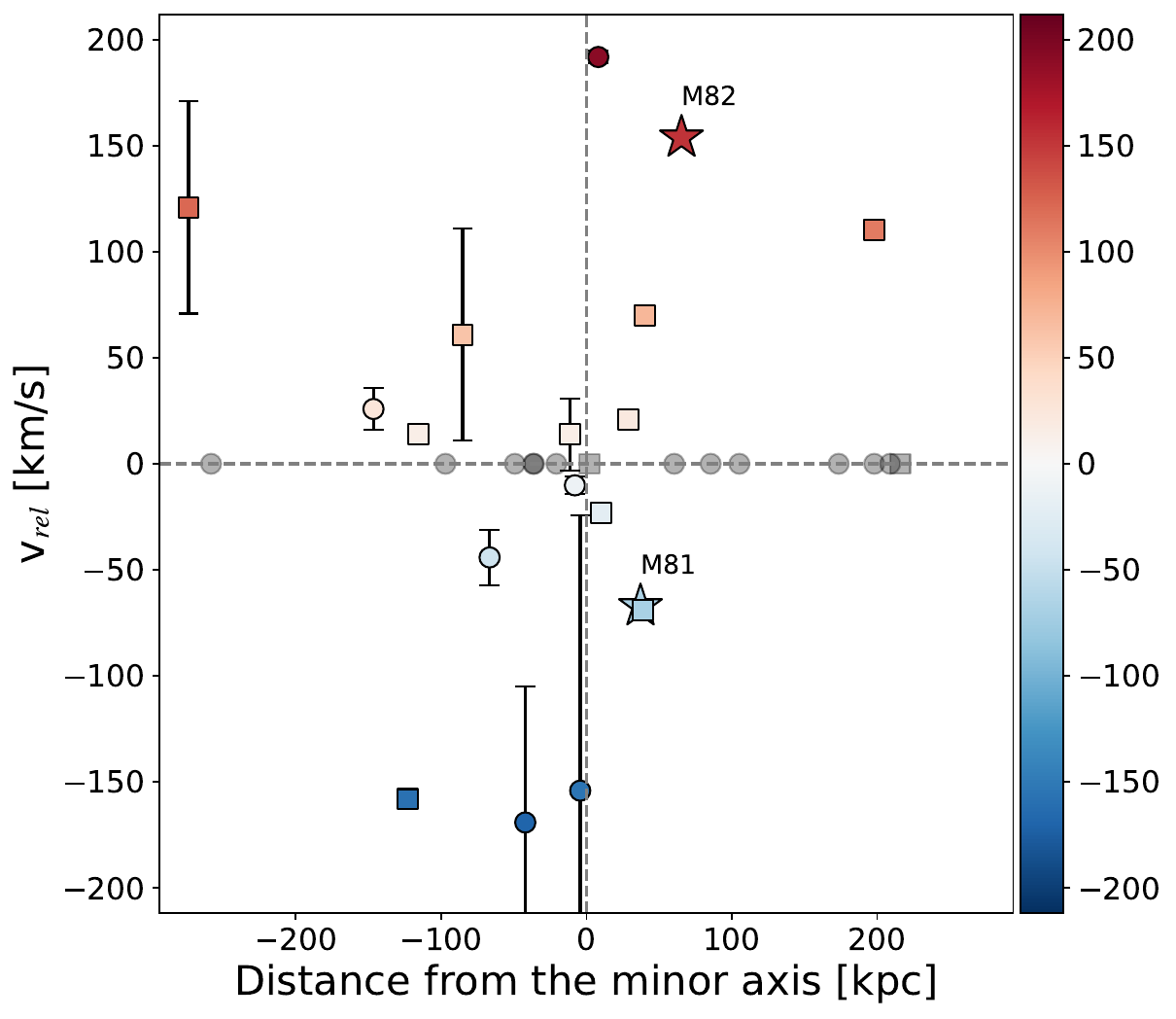}

    \caption{A position-velocity diagram of the satellite system of the M\,81 group. Blue denotes a satellite moving towards us, red away from us, relative to the mean group velocity ($v$=29\,km/s). {The gray symbols indicate the structure members without available line-of-sight velocities.} The x-axis denotes the distance from the minor axis along the major axis. A negative sign on the distance is assigned based on their position on the left (positive x) of the minor axis. The circles highlight the dwarf spheroidals, the squares are dwarfs of all other morphologies. The two stars indicate the two main galaxies M\,81 and M\,82.}
    \label{fig:pv}
\end{figure}

{Velocity measurements are available for 19 dwarf galaxies,  of which 16 are members of the reported flattened structure}. This is comparable to available line-of-sight data for the Andromeda galaxy and Cen\,A. However, these galaxy groups are more simple in terms of their dynamics, because they are both dominated by a single giant galaxy. On the other hand, the M\,81 group consists of at least two main galaxies -- M\,81 and M\,82 -- as well as several massive dwarfs such as NGC\,2976 and NGC\,3077. This makes an analysis of the dynamics of the group not straightforward. For example, M\,81 has a line-of-sight velocity of -38\,km\,s$^{-1}$ \citep{1978ApJ...223..391D} and M\,82 of 183\,km\,s$^{-1}$ \citep{1981MNRAS.195..327A}. The satellites' mean line-of-sight velocity is 20\,km\,s$^{-1}$, which is offset by $\approx$60\,km\,s$^{-1}$ from M\,81's velocity. In comparison, the difference between the velocity of Cen\,A and the mean of its satellite system is 1\,km\,s$^{-1}$ \citep{Muller2021b}. If we take a simple approach and use the mean satellite velocity {of the dwarfs that are found to be part of the structure} (29\,km\,s$^{-1}$) as an anchor which coincides with the luminosity weighted velocity of M\,81/M\,82 (29\,km\,s$^{-1}$), we can produce a position-velocity (PV) diagram for the flattened structure, see Fig.\,\ref{fig:pv}. In general, we expect that a fully co-rotating system would occupy two out of the four quadrants on opposing sides, and a fully pressure-supported system all four quadrants. For the M\,81 system, a clear signal of corotation is not obvious. The two opposing quadrants each populate 9 and 7 satellites, respectively. In this plot, we used the 2D center of the flattened dwarf system as found with the Hough fit as the zero point for the distance. There may be other choices to be made, e.g. either using M\,81 as the center, or a weighted mean between M\,81 and M\,82. Using the stellar luminosities as a proxy for the latter, we find only a minor difference (11\,kpc) to the position of M\,81, which is negligible. Fixing the center at M\,81, most dwarf galaxies would populate the two left quadrants (it would shift the vertical line to the right in Fig\,\ref{fig:pv}). 

Assuming the flattened satellite system of M\,81 is observed edge-on, we can study the line-of-sight velocities of the dwarfs from a face-on point-of-view of the system. Such a plot is presented in Fig.\,\ref{fig:SG}, bottom right panel. The line-of-sight-velocities will be in the plane of this panel. There are a few interesting points to note. Almost all velocities are coming from dwarfs on one side, which coincides with the position of NGC\,3077. And most of these are dIrrs. Could it be that at least a few of these dwarfs were a distinct group including NGC\,3077 and are falling in? Because the HI in dwarf galaxies should get stripped when accreted by a massive galaxy such as  M\,81 \citep{2014ApJ...795L...5S}, it is possible that we are observing their (first) infall. {Moreover, between M\,81 and NGC\,3077, there are five candidate tidal dwarf galaxies (d0959+68, BK3N, Garland, A0952+69, and Holm IX) that are embedded in the HI tidal material, so they may further complicate the dynamical interpretation of the system (see, e.g., \citealt{2015A&A...584A.113L} for bona-fide tidal dwarf galaxies in other interacting systems). Out of these five candidates all but d0959+68 have velocity measurements. Two would be co-moving as expected for a co-rotating plane-of-satellites, two would be off. {Tidal dwarf galaxies may offer a possible formation scenario for the creation of planes-of-satellites, with numerical fly-by models showing that up to 50\% of the tidal dwarf galaxies may end up on counter-rotating orbits \citep{2011A&A...532A.118P}. However, as long as we do not have proper motion measurements for these dwarfs, we cannot assess whether they are truly co/counter-rotating, only that they may be consistent with such a motion (see e.g. \citealt{2023MNRAS.519.6184K}). }}

{With the current incomplete picture of the satellite system (i.e. the missing velocities of half the dwarf galaxies) we cannot draw firm conclusions about the dynamical state of the system. But currently, there is no strong evidence for co-rotation of the satellite system.}

\section{Comparison to Illustris-TNG50}
\label{simulations}

\begin{figure*}[htb]
    \centering
    \includegraphics[width=0.49\linewidth]{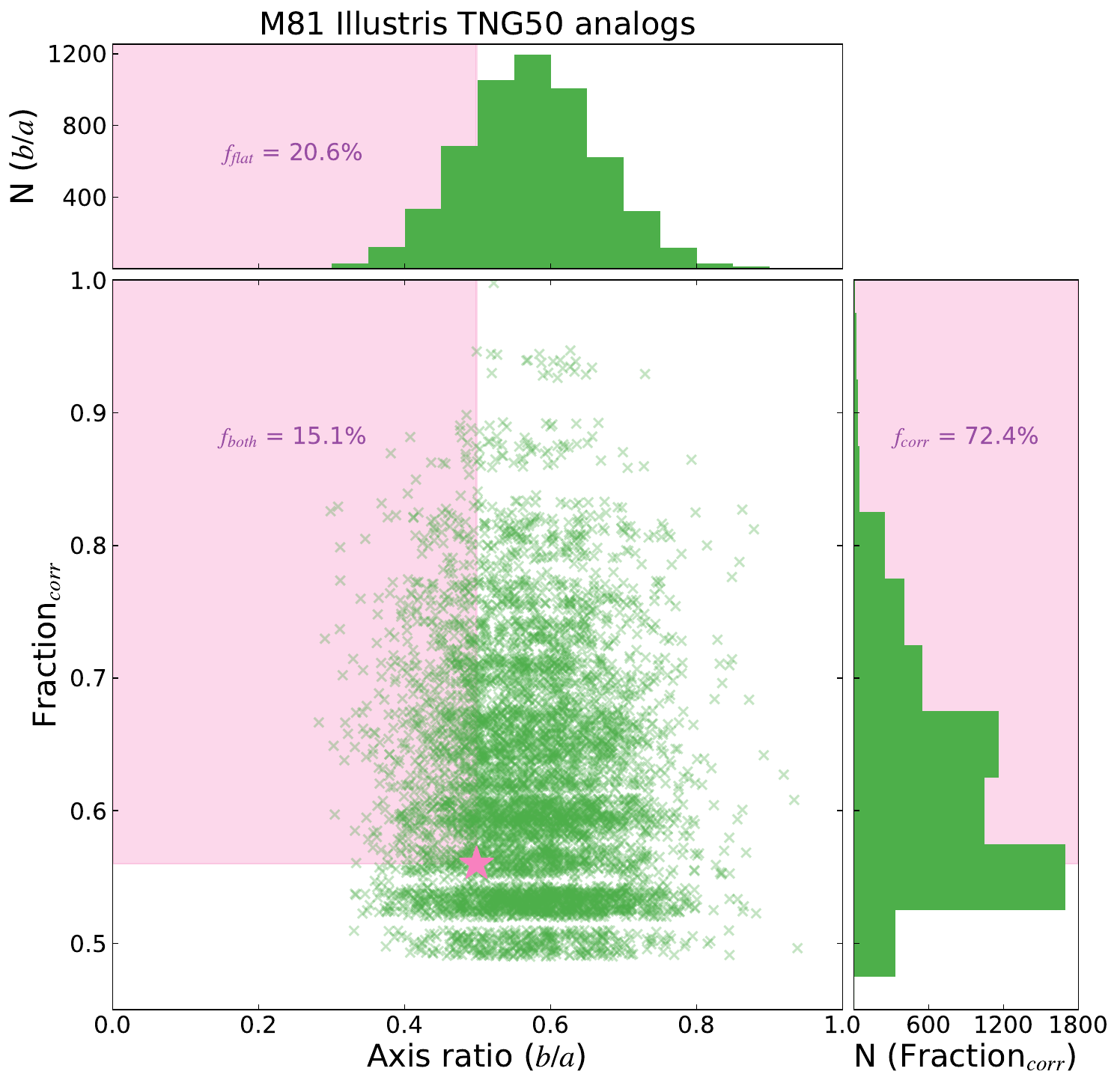}
    \includegraphics[width=0.49\linewidth]{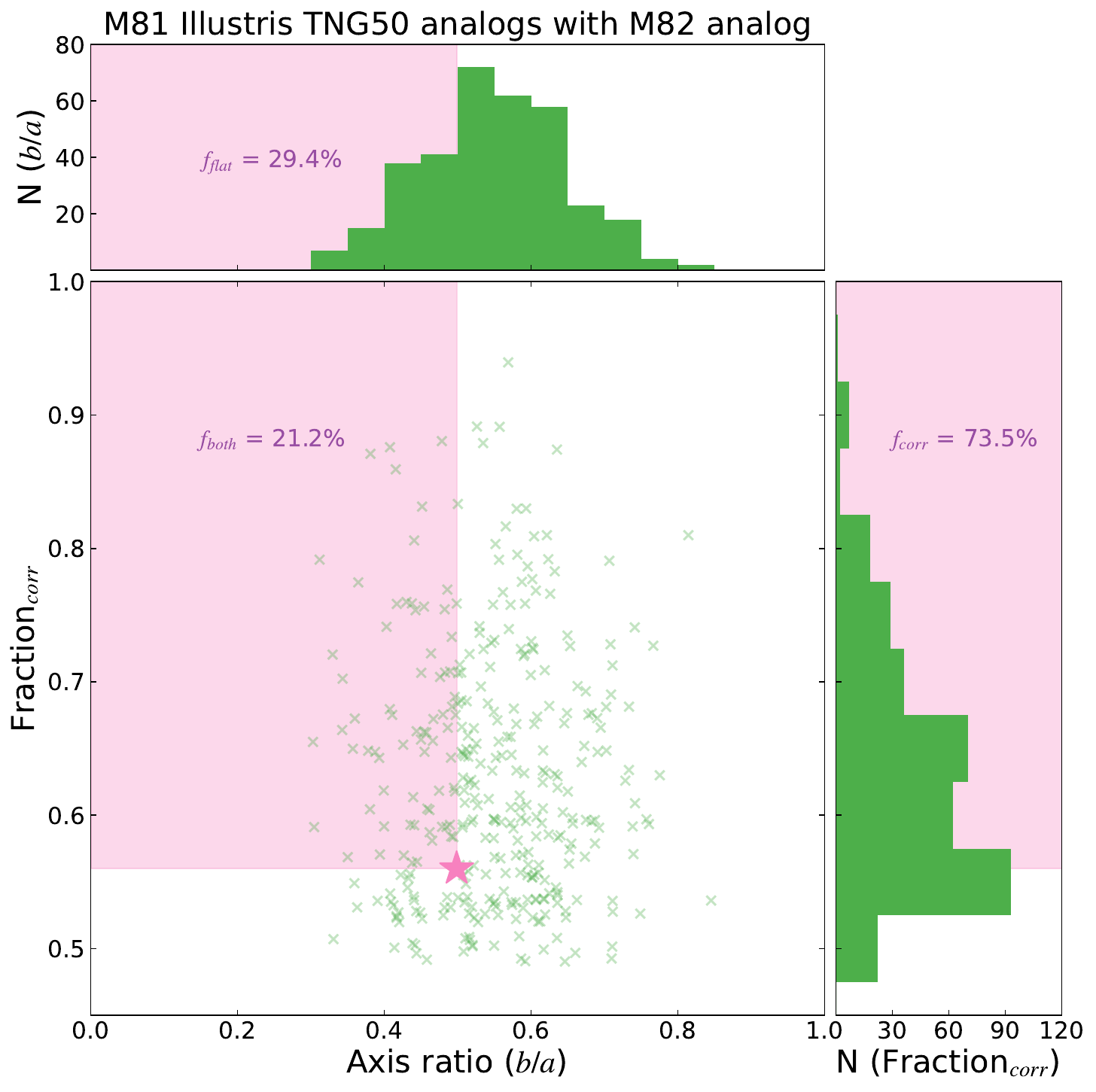}

    \caption{Comparison of the M81 flattened structure parameters with M81-analogs from TNG50. On the x-axis, we plot the short over long axis ratio ($b/a$), and on the y-axis the number of corotating satellites. The green crosses show the parameters of the M81-analogs in simulations while the pink star shows the parameter pair of the data. The top histogram shows the distribution of the axis ratios of the structures around simulated analogs, and the histogram on the right shows the corresponding number of kinematically correlated dwarfs. The pink-shaded regions illustrate the analog systems that match or exceed the parameters found in the data. The inset texts indicate the corresponding frequencies of such systems in simulations. {Left: results for all simulated M81 analogs. Right: results for the sub-sample of analogs that also feature an M82 analog.}}
    \label{fig:sim_comp}
\end{figure*}

To probe the significance of M81's flattened satellite distribution, we use data from the IllustrisTNG suite \citep{2018MNRAS.473.4077P} of publicly available, large-volume hydrodynamic simulations which adopts cosmological parameters from Planck \citep{2016A&A...594A..13P}: $ega_{\Lambda} = 0.6911$, $ega_\mathrm{m}=0.3089$, and $h=0.6774$. To best resolve the fainter dwarf population in the M81 group, we make use of the high-resolution TNG50-1 run -- spanning a simulation volume of length $51.7\,\mathrm{Mpc}$ with dark and gas particle resolutions of $M_{\mathrm{DM}}=4.5\times10^{5}M_{\odot}$ and $M_{\mathrm{gas}}=8.5\times10^{4}M_{\odot}$ respectively.

We define M81's simulated analogs as having a stellar mass between $5\times10^{10}\,M_{\odot}$ and $2\times10^{11}\,M_{\odot}$ while also requiring an associated halo mass $M_{200}$ between $5\times10^{11}\,M_{\odot}$ and $2\times10^{12}\,M_{\odot}$ (as motivated by M81's dynamical mass estimate of $1.17\times10^{12}\,M_{\odot}$ from \citealt{2017MNRAS.467..273O}). Analogs must also be isolated from other dwarf associations -- specifically, we reject all systems with a companion halo $250-1000\,\mathrm{kpc}$ away with a halo mass above $0.25M_{200}$. {This gives us our sample of M\,81 analogs.} Additionally, we inspect each M81 system analog for a corresponding M82 analog, defined as a subhalo within $200\,\mathrm{kpc}$ of the host galaxy with a stellar mass above $0.25\,M_{*,\mathrm{M81}}$ -- thus accommodating the M81-M82 stellar mass ratio of $\sim0.43$ \citep{2013AJ....145..101K}. {In the first step we analyze the full analog sample. In the second step, we restrict the sample by only considering systems that also feature an M82 analog and repeat our analysis.}

Each M81 analog is mock-observed from {10} isotropically distributed directions with corresponding distances drawn from M81's distance modulus of $\mu=27.84\pm0.09$. Since M81's satellite population is complete within a projected $230\,\mathrm{kpc}$ radius to a limiting magnitude of $-$10 in the $r$ band, we identify all subhaloes within a projected radius of $250\,\mathrm{kpc}$ with a depth of $500\,\mathrm{kpc}$ with respect to their host galaxy. The subhaloes are ranked by their stellar mass (then dark mass once no stellar particles are available), and the 34 most massive are taken as the realization's satellite sample. Finally, distance errors are drawn from M81's observed satellites and randomly associated to the simulated satellites along the mock-observed line-of-sight. 

\begin{figure}[htb]
    \centering
    \includegraphics[width=\linewidth]{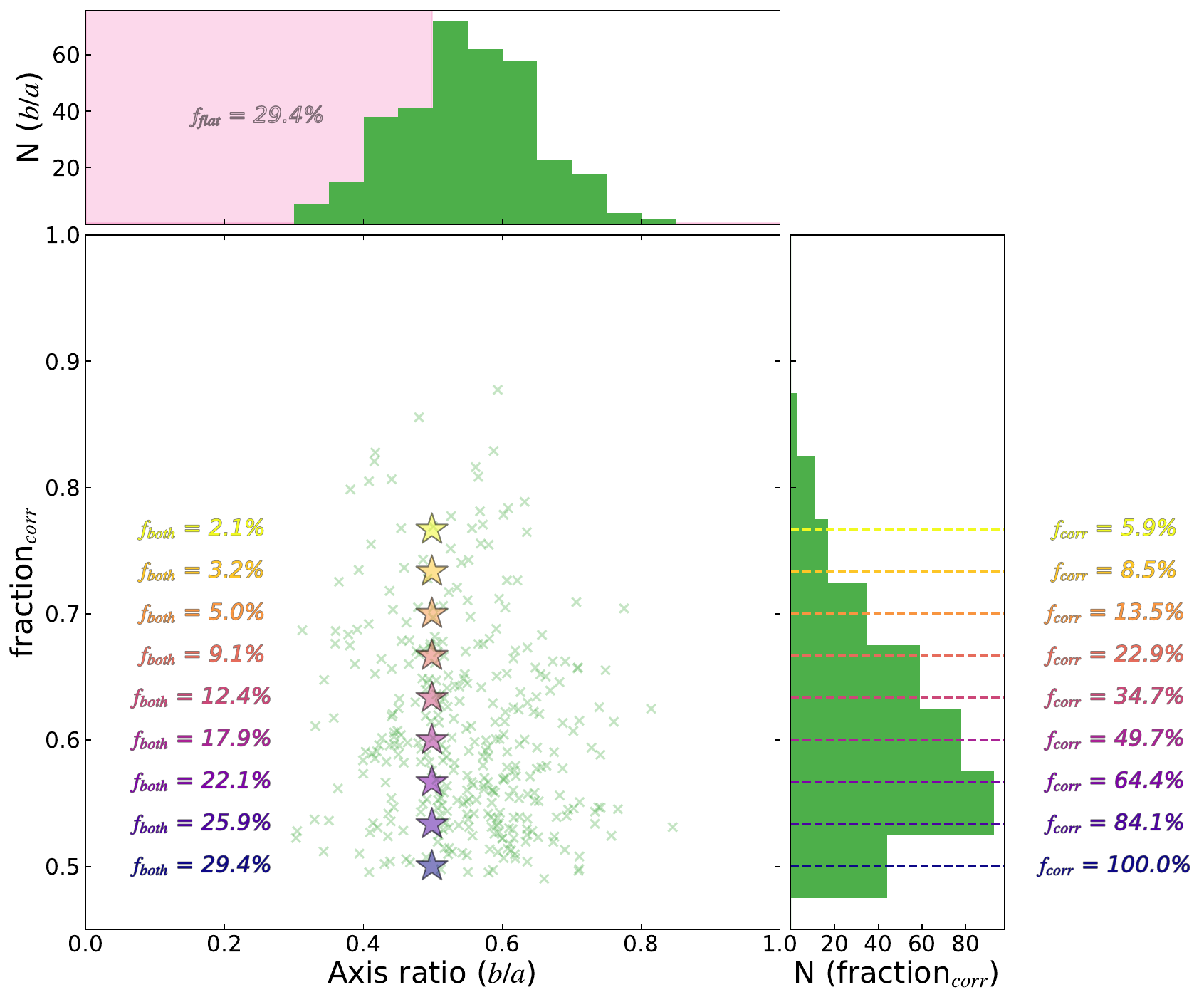}

    \caption{{Same as Figure \ref{fig:sim_comp} (right) but illustrating the possible outcomes given missing velocity measurements. The colored stars show the possible axis ratios and fractions of corotation in the data. The flatness of the structure will not change with new velocity information because distance measurements are available for all dwarfs. The fractions of corotation can vary from 0.5 (15/30) to 0.76 (23/30). For each possible scenario the fraction of simulated systems which match or exceed the degree of kinematic correlation is shown on the histogram to the right. The fractions of systems which show both flatness and degree of corotation as or more extreme as the data are shown in the center.}}
    \label{fig:sim_comp_pred}
\end{figure}

{In order to assess the prevalence of similarly flattened structures in simulations, we perform the described Hough fitting procedure on the 5530 {(340 featuring an M82 analog)} 
M81 analog systems.  For this, we use the same parameter search area which was found to optimize the flattening and the number of structure members in the observed M81 satellite system. We compare both the flattening and the degree of kinematic correlation, i.e., the number of corotating satellites. {The Hough fitting rejects 0 to 12 outliers (compared to four for the observations). {The mean number of rejected outliers is 4 (as is the number of observed rejected outliers) with a standard deviation of 1.8.} We have looked into a potential correlation between the number of outliers and the axis ratio of the resulting structures. In general we would expect that having a smaller sample (i.e. more outliers) would lead to a flatter structure \citep{2017ApJ...850..132P}. However, there is no such correlation when we compare the distributions of b/a ratios for the different samples with outliers between 0 to 12. This is due to how we set-up the search radius during the Hough fitting on the simulated systems. We require it to be the same as for the observed system (maximizing flatness and number of objects considered in the fitting). If we would not require this, we would indeed find a correlation between the thickness and the sample size.
%We do not distinguish for the measurement of the b/a ratios with the Hough fit between the samples with a different number of outliers.
} 

Because not all satellites have available velocity measurements, we assess {the fraction of observed structure members that has such estimates (53\%). When finding a linear substructure among outliers, the Hough fitting technique is not forced to include the same number of satellites in the simulations that are found to be part of the structure (30). As discussed before, the method selects a variable number of dwarfs, which is very similar to the number of structure members in observations. Because the structure populations are slightly variable in every simulated analog system, we select 53\% of the velocities from the structure member satellites in the simulated systems. We use the satellites with the highest dark matter mass (as a proxy for their stellar mass) for this and use these to determine the fraction of corotating satellites along the major axis of a given structure.} 
% how many structure members in the data have such estimates and correspondingly select the velocities of the 16 satellites with the highest dark matter mass (as a proxy for their stellar mass) in the simulation. We use these to determine the number of corotating satellites along the major axis of a given structure. 
Similarly to the data analysis, we use the mean velocity of these satellites and the minor axis of their spatial distribution as anchor points for their corotation. The described comparison results in three measures of significance or p-values: the frequency of systems that match or exceed the observed axis radio, the degree of kinematic correlation, and the simultaneous occurrence of both properties. We find the corresponding p-values {for the full sample and the sample featuring an M82 analog, respectively}: $p_{flat}$ = {0.206 (0.294)}
, $p_{kin}$ = {0.724 (0.735)}
and $p_{both}$ = {0.151 (0.212)}.
 We thus estimate that the observed structure occurs in approximately {15.1\% (21.2\%)}
of M81-like systems. %The behavior of this structure is therefore consistent with cosmological expectations. 
{The results for both samples are illustrated in Figure \ref{fig:sim_comp}.} 
%We show the axis ratio ($b/a$) of the structures found via the Hough transform vs. the number of kinematically correlated satellites in these structures. The parameter pair observed in the M81 structure is shown in comparison with results from simulation analogs. The parameter locations of simulated analogs that match or exceed the reported structure's flatness, kinematic correlation, and both of these simultaneously are highlighted. Two histograms illustrate the distribution of structure parameters found in the analog systems.

{Since not all M81 satellites have available velocity measurements at this time, the degree of corotation in the structure members is still unknown. Inspecting the structure members we determine that the minimum number of possible corotating structure members is 15 (50\%) and the maximum number is 23 (76.6\%). In a second comparison between the data and the simulated systems, we consider the velocities of all structure members in simulations to determine the fraction of corotation. We study all possible outcomes in terms of corotation given the missing velocities in the data and compare all possibilities with the expectations from the simulated analog systems. This results in a p-value for every possible fraction of corotation and allows us to make a prediction on how many satellites are expected to corotate in order to be consistent with $\Lambda$CDM. The results of this comparison are illustrated in Figure \ref{fig:sim_comp_pred}. The fractions of simulated systems for which both the flatness and degree of corotation match or exceed the possible values with future observations range from 2.1\% to 29.4\%. This illustrates that the consistency with the expectations from cosmological simulations is still uncertain due to the missing velocities and therefore unknown degree of corotation in the reported structure.} 
}

\section{Summary and conclusions}
The plane-of-satellites problem has been identified as one of the biggest challenges to cosmology on small scales \citep{2022NatAs...6..897S}. Therefore, it is imperative to study different galaxy groups and test whether the motion and distribution of satellite galaxies follows predictions from cosmological simulations or not. In this work, we quantified previous claims of a flattened structure of dwarf galaxies in the compact M\,81 group. We find that 30 out of 34 dwarf galaxies follow a flattened structure with a minor-to-major axis ratio of $b/a = 0.5$, which is similar to that of Cen\,A \citep{Muller2019} and of early-type galaxies in the nearby universe (10-45\,Mpc, \citealt{2021A&A...654A.161H}) from the MATLAS survey \citep{2020MNRAS.491.1901H,2021MNRAS.506.5494P}. If we consider only the dwarf spheroidals, the flattening increases to a ratio of 0.34. This is still thicker than the best studied planes around the Milky Way and the Andromeda galaxy (c/a\,$<$\,0.2). However, for the Milky Way and Andromeda systems, accurate distance estimates are available, so the axis ratio can be estimated in 3D. Here, the thickness was measured in 2D. If the planar structure is significantly inclined with respect to the line of sight, the projected 2D thickness will increase. The difference between the thickness of the dSph-only and the dSph+dIrr samples may also arise from the lower numbers of tracers. The fewer tracers, the more likely it is to get a thinner spatial distribution \citep{2017ApJ...850..132P}. 
%When comparing with cosmological simulations, this fact must be considered, otherwise one may bias the results.

How consistent is the flattened structure with cosmological predictions? We have compared the on-sky flattening and motion of the satellites in the M\,81 group with TNG50 from the IllustrisTNG suite. In contrast to the Local Group and Centaurus system, we find agreement within 2$\sigma$ with $\Lambda$CDM expectations. This is an interesting result because M\,81 is in a different dynamical state than the Milky Way with its quiet merger history, and Andromeda and Cen\,A with putative major mergers 2-3\,Gyr ago \citep{2018MNRAS.475.2754H,2020MNRAS.498.2766W}.

Considering the possible coherent motion of satellites, which is the crucial property of satellite systems to test $\Lambda$CDM, we find no correlation for the M\,81 satellites. This is in contrast to the Milky Way, the M\,31, and Cen\,A satellites, which all seem to follow a coherent motion around their hosts. Why could this be?

The M\,81 group is a compact group which is currently undergoing a major merger and is therefore not in  dynamical equilibrium. This is evident from strong signs of interactions between M\,81, M\,82, and NGC3077, especially in HI \citep{2008AJ....135.1983C}. Even if there was a co-rotating plane-of-satellites around M\,81 before, such an encounter may destroy any signs of co-rotation. Simulating the effect of major mergers on the creation or destruction of planes-of-satellites could help understand the observations of the M\,81 group. In this respect, simulations of the major merger of Cen\,A hints toward a connection between the merger and the plane-of-satellites \citep{2020MNRAS.498.2766W}, but more detailed simulations are needed to reproduce the Cen\,A system. However, studies using TNG100 from the IllustrisTNG suite of cosmological simulations show a negligible contribution of the major merger history on the phase-space coherence of satellite galaxies \citep{2023MNRAS.524..952K}. Under this context, we would neither expect the formation or destruction of a co-rotating satellite system around M\,81.

Another explanation could be that we are seeing a mixing of several dwarf galaxy systems. It may be that a group of dwarf galaxies is falling in together. This is hinted at by the clustered distribution of dIrrs towards one side of M\,81. {In addition, at least 4 out of 19 satellites for which we have velocity information (d0952+69, BK3N, HolmIX, and Garland) are candidate tidal dwarf galaxies, being embedded in the HI tidal material between M\,81 and NGC\,3077. Two of those are reducing the phase-space signal.}
Before ruling out a coherent motion of satellites, it is imperative to sample the full dwarf galaxy population around M\,81, especially dwarf spheroidal galaxies for which we can rule out a recent infall or a recent tidal formation. There are 15 more dwarf galaxies awaiting velocity measurements, and many of those are dSphs, for which we find a more significant flattening than for the whole population. {By studying the possible degrees of corotation in the reported flattened structure in light of the missing velocity measurements, we can make a prediction on the expected fraction of corotating satellites such that the observed structure is consistent with its simulated analogs. The fractions of simulated analogs which match or exceed both the observed flatness and kinematic correlation range from 2.1\% to 29.4\%. This indicates that the consistency with expectations from state-of-the-art cosmological simulations is still uncertain until the missing 15 velocity measurements are obtained.}

%We refrain from making a comparison to cosmological simulations in this work. The reason is that the M\,81 group is a complicated system with many constraints. For example, fixing the separation and velocities between M\,81 and M\,82 (and potentially that of NGC\,3077 and IC\,2574), as well as their masses, will likely leave only few analogs in cosmological simulations. Then drawing conclusions about the agreement/disagreement with cosmology from the satellite systems these few analogs will be based on the small number of analogs and may bias the result.

\label{summary}

\begin{acknowledgements} 
{We thank the referee for the constructive report, which helped to clarify and improve the manuscript.}
O.M. and N.H. are grateful to the Swiss National Science Foundation for financial support under the grant number  	PZ00P2\_202104. 
M.S.P. acknowledges funding of a Leibniz-Junior Research Group (project number J94/2020) 
and a KT Boost Fund by the German Scholars Organization and Klaus Tschira Stiftung.
\end{acknowledgements}

\bibliographystyle{aa}
\bibliography{bibliographie}

\begin{thebibliography}{82}
\expandafter\ifx\csname natexlab\endcsname\relax\def\natexlab#1{#1}\fi

\bibitem[{{Appleton} {et~al.}(1981){Appleton}, {Davies}, \&
  {Stephenson}}]{1981MNRAS.195..327A}
{Appleton}, P.~N., {Davies}, R.~D., \& {Stephenson}, R.~J. 1981, \mnras, 195,
  327

\bibitem[{{Bell} {et~al.}(2022){Bell}, {Smercina}, {Price}, {D'Souza},
  {Bailin}, {de Jong}, {Gozman}, {Jang}, {Monachesi}, {Gnedin}, \&
  {Slater}}]{2022ApJ...937L...3B}
{Bell}, E.~F., {Smercina}, A., {Price}, P.~A., {et~al.} 2022, \apjl, 937, L3

\bibitem[{{Cautun} {et~al.}(2015){Cautun}, {Bose}, {Frenk}, {Guo}, {Han},
  {Hellwing}, {Sawala}, \& {Wang}}]{2015MNRAS.452.3838C}
{Cautun}, M., {Bose}, S., {Frenk}, C.~S., {et~al.} 2015, \mnras, 452, 3838

\bibitem[{{Chiboucas} {et~al.}(2013){Chiboucas}, {Jacobs}, {Tully}, \&
  {Karachentsev}}]{2013AJ....146..126C}
{Chiboucas}, K., {Jacobs}, B.~A., {Tully}, R.~B., \& {Karachentsev}, I.~D.
  2013, \aj, 146, 126

\bibitem[{{Chiboucas} {et~al.}(2009){Chiboucas}, {Karachentsev}, \&
  {Tully}}]{2009AJ....137.3009C}
{Chiboucas}, K., {Karachentsev}, I.~D., \& {Tully}, R.~B. 2009, \aj, 137, 3009

\bibitem[{{Chynoweth} {et~al.}(2008){Chynoweth}, {Langston}, {Yun}, {Lockman},
  {Rubin}, \& {Scoles}}]{2008AJ....135.1983C}
{Chynoweth}, K.~M., {Langston}, G.~I., {Yun}, M.~S., {et~al.} 2008, \aj, 135,
  1983

\bibitem[{{Conn} {et~al.}(2013){Conn}, {Lewis}, {Ibata}, {Parker}, {Zucker},
  {McConnachie}, {Martin}, {Valls-Gabaud}, {Tanvir}, {Irwin}, {Ferguson}, \&
  {Chapman}}]{2013ApJ...766..120C}
{Conn}, A.~R., {Lewis}, G.~F., {Ibata}, R.~A., {et~al.} 2013, \apj, 766, 120

\bibitem[{{Dalcanton} {et~al.}(2009){Dalcanton}, {Williams}, {Seth}, {Dolphin},
  {Holtzman}, {Rosema}, {Skillman}, {Cole}, {Girardi}, {Gogarten},
  {Karachentsev}, {Olsen}, {Weisz}, {Christensen}, {Freeman}, {Gilbert},
  {Gallart}, {Harris}, {Hodge}, {de Jong}, {Karachentseva}, {Mateo}, {Stetson},
  {Tavarez}, {Zaritsky}, {Governato}, \& {Quinn}}]{2009ApJS..183...67D}
{Dalcanton}, J.~J., {Williams}, B.~F., {Seth}, A.~C., {et~al.} 2009, \apjs,
  183, 67

\bibitem[{{de Vaucouleurs}(1956)}]{1956VA......2.1584D}
{de Vaucouleurs}, G. 1956, Vistas in Astronomy, 2, 1584

\bibitem[{{Dickel} \& {Rood}(1978)}]{1978ApJ...223..391D}
{Dickel}, J.~R. \& {Rood}, H.~J. 1978, \apj, 223, 391

\bibitem[{{Ferrarese} {et~al.}(2000){Ferrarese}, {Ford}, {Huchra}, {Kennicutt},
  {Mould}, {Sakai}, {Freedman}, {Stetson}, {Madore}, {Gibson}, {Graham},
  {Hughes}, {Illingworth}, {Kelson}, {Macri}, {Sebo}, \&
  {Silbermann}}]{2000ApJS..128..431F}
{Ferrarese}, L., {Ford}, H.~C., {Huchra}, J., {et~al.} 2000, \apjs, 128, 431

\bibitem[{{Habas} {et~al.}(2020){Habas}, {Marleau}, {Duc}, {Durrell}, {Paudel},
  {Poulain}, {S{\'a}nchez-Janssen}, {Sreejith}, {Ramasawmy}, {Stemock},
  {Leach}, {Cuillandre}, {Gwyn}, {Agnello}, {B{\'\i}lek}, {Fensch},
  {M{\"u}ller}, {Peng}, \& {van der Burg}}]{2020MNRAS.491.1901H}
{Habas}, R., {Marleau}, F.~R., {Duc}, P.-A., {et~al.} 2020, \mnras, 491, 1901

\bibitem[{{Hammer} {et~al.}(2018){Hammer}, {Yang}, {Wang}, {Ibata}, {Flores},
  \& {Puech}}]{2018MNRAS.475.2754H}
{Hammer}, F., {Yang}, Y.~B., {Wang}, J.~L., {et~al.} 2018, \mnras, 475, 2754

\bibitem[{{Heesters} {et~al.}(2021){Heesters}, {Habas}, {Marleau},
  {M{\"u}ller}, {Duc}, {Poulain}, {Durrell}, {S{\'a}nchez-Janssen}, \&
  {Paudel}}]{2021A&A...654A.161H}
{Heesters}, N., {Habas}, R., {Marleau}, F.~R., {et~al.} 2021, \aap, 654, A161

\bibitem[{{Hoffman} {et~al.}(2018){Hoffman}, {Carlesi}, {Pomar{\`e}de},
  {Tully}, {Courtois}, {Gottl{\"o}ber}, {Libeskind}, {Sorce}, \&
  {Yepes}}]{2018NatAs...2..680H}
{Hoffman}, Y., {Carlesi}, E., {Pomar{\`e}de}, D., {et~al.} 2018, Nature
  Astronomy, 2, 680

\bibitem[{{Hoffman} {et~al.}(2012){Hoffman}, {Metuki}, {Yepes},
  {Gottl{\"o}ber}, {Forero-Romero}, {Libeskind}, \&
  {Knebe}}]{2012MNRAS.425.2049H}
{Hoffman}, Y., {Metuki}, O., {Yepes}, G., {et~al.} 2012, \mnras, 425, 2049

\bibitem[{Hough(1959)}]{hough1959machine}
Hough, P.~V. 1959, in Conf. Proc., Vol. 590914, 554--558

\bibitem[{Hough(1962)}]{hough1962method}
Hough, P.~V. 1962, Method and means for recognizing complex patterns, uS Patent
  3,069,654

\bibitem[{{Huchtmeier} {et~al.}(2003){Huchtmeier}, {Karachentsev}, \&
  {Karachentseva}}]{2003A&A...401..483H}
{Huchtmeier}, W.~K., {Karachentsev}, I.~D., \& {Karachentseva}, V.~E. 2003,
  \aap, 401, 483

\bibitem[{{Ibata} {et~al.}(2014{\natexlab{a}}){Ibata}, {Ibata}, {Famaey}, \&
  {Lewis}}]{2014Natur.511..563I}
{Ibata}, N.~G., {Ibata}, R.~A., {Famaey}, B., \& {Lewis}, G.~F.
  2014{\natexlab{a}}, \nat, 511, 563

\bibitem[{{Ibata} {et~al.}(2014{\natexlab{b}}){Ibata}, {Ibata}, {Lewis},
  {Martin}, {Conn}, {Elahi}, {Arias}, \& {Fernando}}]{2014ApJ...784L...6I}
{Ibata}, R.~A., {Ibata}, N.~G., {Lewis}, G.~F., {et~al.} 2014{\natexlab{b}},
  \apjl, 784, L6

\bibitem[{{Ibata} {et~al.}(2013){Ibata}, {Lewis}, {Conn}, {Irwin},
  {McConnachie}, {Chapman}, {Collins}, {Fardal}, {Ferguson}, {Ibata}, {Mackey},
  {Martin}, {Navarro}, {Rich}, {Valls-Gabaud}, \&
  {Widrow}}]{2013Natur.493...62I}
{Ibata}, R.~A., {Lewis}, G.~F., {Conn}, A.~R., {et~al.} 2013, NAT, 493, 62

\bibitem[{{Kanehisa} {et~al.}(2023{\natexlab{a}}){Kanehisa}, {Pawlowski}, \&
  {M{\"u}ller}}]{2023MNRAS.524..952K}
{Kanehisa}, K.~J., {Pawlowski}, M.~S., \& {M{\"u}ller}, O. 2023{\natexlab{a}},
  \mnras, 524, 952

\bibitem[{{Kanehisa} {et~al.}(2023{\natexlab{b}}){Kanehisa}, {Pawlowski},
  {M{\"u}ller}, \& {Sohn}}]{2023MNRAS.519.6184K}
{Kanehisa}, K.~J., {Pawlowski}, M.~S., {M{\"u}ller}, O., \& {Sohn}, S.~T.
  2023{\natexlab{b}}, \mnras, 519, 6184

\bibitem[{{Karachentsev} {et~al.}(2006){Karachentsev}, {Dolphin}, {Tully},
  {Sharina}, {Makarova}, {Makarov}, {Karachentseva}, {Sakai}, \&
  {Shaya}}]{2006AJ....131.1361K}
{Karachentsev}, I.~D., {Dolphin}, A., {Tully}, R.~B., {et~al.} 2006, \aj, 131,
  1361

\bibitem[{{Karachentsev} {et~al.}(2002){Karachentsev}, {Dolphin}, {Geisler},
  {Grebel}, {Guhathakurta}, {Hodge}, {Karachentseva}, {Sarajedini}, {Seitzer},
  \& {Sharina}}]{2002A&A...383..125K}
{Karachentsev}, I.~D., {Dolphin}, A.~E., {Geisler}, D., {et~al.} 2002, \aap,
  383, 125

\bibitem[{{Karachentsev} {et~al.}(1985){Karachentsev}, {Karachentseva}, \&
  {Boerngen}}]{1985MNRAS.217..731K}
{Karachentsev}, I.~D., {Karachentseva}, V.~E., \& {Boerngen}, F. 1985, \mnras,
  217, 731

\bibitem[{{Karachentsev} {et~al.}(2000){Karachentsev}, {Karachentseva},
  {Dolphin}, {Geisler}, {Grebel}, {Guhathakurta}, {Hodge}, {Sarajedini},
  {Seitzer}, \& {Sharina}}]{2000A&A...363..117K}
{Karachentsev}, I.~D., {Karachentseva}, V.~E., {Dolphin}, A.~E., {et~al.} 2000,
  \aap, 363, 117

\bibitem[{{Karachentsev} {et~al.}(2004){Karachentsev}, {Karachentseva},
  {Huchtmeier}, \& {Makarov}}]{2004AJ....127.2031K}
{Karachentsev}, I.~D., {Karachentseva}, V.~E., {Huchtmeier}, W.~K., \&
  {Makarov}, D.~I. 2004, \aj, 127, 2031

\bibitem[{{Karachentsev} \& {Kashibadze}(2005)}]{2005astro.ph..9207K}
{Karachentsev}, I.~D. \& {Kashibadze}, O.~G. 2005, arXiv e-prints, astro

\bibitem[{{Karachentsev} {et~al.}(2013){Karachentsev}, {Makarov}, \&
  {Kaisina}}]{2013AJ....145..101K}
{Karachentsev}, I.~D., {Makarov}, D.~I., \& {Kaisina}, E.~I. 2013, \aj, 145,
  101

\bibitem[{{Karachentsev} {et~al.}(2003){Karachentsev}, {Makarov}, {Sharina},
  {Dolphin}, {Grebel}, {Geisler}, {Guhathakurta}, {Hodge}, {Karachentseva},
  {Sarajedini}, \& {Seitzer}}]{2003A&A...398..479K}
{Karachentsev}, I.~D., {Makarov}, D.~I., {Sharina}, M.~E., {et~al.} 2003, \aap,
  398, 479

\bibitem[{{Karachentsev} {et~al.}(2001){Karachentsev}, {Sharina}, {Dolphin},
  {Geisler}, {Grebel}, {Guhathakurta}, {Hodge}, {Karachentseva}, {Sarajedini},
  \& {Seitzer}}]{2001A&A...375..359K}
{Karachentsev}, I.~D., {Sharina}, M.~E., {Dolphin}, A.~E., {et~al.} 2001, \aap,
  375, 359

\bibitem[{{Koch} \& {Grebel}(2006)}]{2006AJ....131.1405K}
{Koch}, A. \& {Grebel}, E.~K. 2006, \aj, 131, 1405

\bibitem[{{Kroupa} {et~al.}(2005){Kroupa}, {Theis}, \&
  {Boily}}]{2005A&A...431..517K}
{Kroupa}, P., {Theis}, C., \& {Boily}, C.~M. 2005, \aap, 431, 517

\bibitem[{{Lahav} {et~al.}(1998){Lahav}, {Santiago}, {Webster}, {Strauss},
  {Davis}, {Dressler}, \& {Huchra}}]{1998astro.ph..9343L}
{Lahav}, O., {Santiago}, B.~X., {Webster}, A.~M., {et~al.} 1998, arXiv
  e-prints, astro

\bibitem[{{Lang} {et~al.}(2003){Lang}, {Boyce}, {Kilborn}, {Minchin}, {Disney},
  {Jordan}, {Grossi}, {Garcia}, {Freeman}, {Phillipps}, \&
  {Wright}}]{2003MNRAS.342..738L}
{Lang}, R.~H., {Boyce}, P.~J., {Kilborn}, V.~A., {et~al.} 2003, \mnras, 342,
  738

\bibitem[{{Lelli} {et~al.}(2015){Lelli}, {Duc}, {Brinks}, {Bournaud},
  {McGaugh}, {Lisenfeld}, {Weilbacher}, {Boquien}, {Revaz}, {Braine},
  {Koribalski}, \& {Belles}}]{2015A&A...584A.113L}
{Lelli}, F., {Duc}, P.-A., {Brinks}, E., {et~al.} 2015, \aap, 584, A113

\bibitem[{{Libeskind} {et~al.}(2019){Libeskind}, {Carlesi}, {M{\"u}ller},
  {Pawlowski}, {Hoffman}, {Pomar{\`e}de}, {Courtois}, {Tully}, {Gottl{\"o}ber},
  {Steinmetz}, {Sorce}, \& {Knebe}}]{2019MNRAS.490.3786L}
{Libeskind}, N.~I., {Carlesi}, E., {M{\"u}ller}, O., {et~al.} 2019, \mnras,
  490, 3786

\bibitem[{{Libeskind} {et~al.}(2007){Libeskind}, {Cole}, {Frenk}, {Okamoto}, \&
  {Jenkins}}]{2007MNRAS.374...16L}
{Libeskind}, N.~I., {Cole}, S., {Frenk}, C.~S., {Okamoto}, T., \& {Jenkins}, A.
  2007, \mnras, 374, 16

\bibitem[{{Libeskind} {et~al.}(2005){Libeskind}, {Frenk}, {Cole}, {Helly},
  {Jenkins}, {Navarro}, \& {Power}}]{2005MNRAS.363..146L}
{Libeskind}, N.~I., {Frenk}, C.~S., {Cole}, S., {et~al.} 2005, \mnras, 363, 146

\bibitem[{{Libeskind} {et~al.}(2016){Libeskind}, {Guo}, {Tempel}, \&
  {Ibata}}]{2016ApJ...830..121L}
{Libeskind}, N.~I., {Guo}, Q., {Tempel}, E., \& {Ibata}, R. 2016, \apj, 830,
  121

\bibitem[{{Libeskind} {et~al.}(2015){Libeskind}, {Hoffman}, {Tully},
  {Courtois}, {Pomar{\`e}de}, {Gottl{\"o}ber}, \&
  {Steinmetz}}]{2015MNRAS.452.1052L}
{Libeskind}, N.~I., {Hoffman}, Y., {Tully}, R.~B., {et~al.} 2015, \mnras, 452,
  1052

\bibitem[{{Libeskind} {et~al.}(2018){Libeskind}, {van de Weygaert}, {Cautun},
  {Falck}, {Tempel}, {Abel}, {Alpaslan}, {Arag{\'o}n-Calvo}, {Forero-Romero},
  {Gonzalez}, {Gottl{\"o}ber}, {Hahn}, {Hellwing}, {Hoffman}, {Jones},
  {Kitaura}, {Knebe}, {Manti}, {Neyrinck}, {Nuza}, {Padilla}, {Platen},
  {Ramachandra}, {Robotham}, {Saar}, {Shandarin}, {Steinmetz}, {Stoica},
  {Sousbie}, \& {Yepes}}]{2018MNRAS.473.1195L}
{Libeskind}, N.~I., {van de Weygaert}, R., {Cautun}, M., {et~al.} 2018, \mnras,
  473, 1195

\bibitem[{{Lynden-Bell}(1976)}]{1976MNRAS.174..695L}
{Lynden-Bell}, D. 1976, \mnras, 174, 695

\bibitem[{{Makarov} {et~al.}(2003){Makarov}, {Karachentsev}, \&
  {Burenkov}}]{2003A&A...405..951M}
{Makarov}, D.~I., {Karachentsev}, I.~D., \& {Burenkov}, A.~N. 2003, \aap, 405,
  951

\bibitem[{{Makarova} {et~al.}(2010){Makarova}, {Koleva}, {Makarov}, \&
  {Prugniel}}]{2010MNRAS.406.1152M}
{Makarova}, L., {Koleva}, M., {Makarov}, D., \& {Prugniel}, P. 2010, \mnras,
  406, 1152

\bibitem[{{Mart{\'\i}nez-Delgado} {et~al.}(2021){Mart{\'\i}nez-Delgado},
  {Makarov}, {Javanmardi}, {Pawlowski}, {Makarova}, {Donatiello}, {Lang},
  {Rom{\'a}n}, {Vivas}, \& {Carballo-Bello}}]{2021A&A...652A..48M}
{Mart{\'\i}nez-Delgado}, D., {Makarov}, D., {Javanmardi}, B., {et~al.} 2021,
  \aap, 652, A48

\bibitem[{{McConnachie} \& {Irwin}(2006)}]{2006MNRAS.365..902M}
{McConnachie}, A.~W. \& {Irwin}, M.~J. 2006, \mnras, 365, 902

\bibitem[{{Metz} {et~al.}(2008){Metz}, {Kroupa}, \&
  {Libeskind}}]{2008ApJ...680..287M}
{Metz}, M., {Kroupa}, P., \& {Libeskind}, N.~I. 2008, \apj, 680, 287

\bibitem[{{M{\"u}ller} {et~al.}(2016){M{\"u}ller}, {Jerjen}, {Pawlowski}, \&
  {Binggeli}}]{Muller2016}
{M{\"u}ller}, O., {Jerjen}, H., {Pawlowski}, M.~S., \& {Binggeli}, B. 2016,
  \aap, 595, A119

\bibitem[{{M{\"u}ller} {et~al.}(2018{\natexlab{a}}){M{\"u}ller}, {Pawlowski},
  {Jerjen}, \& {Lelli}}]{2018Sci...359..534M}
{M{\"u}ller}, O., {Pawlowski}, M.~S., {Jerjen}, H., \& {Lelli}, F.
  2018{\natexlab{a}}, Science, 359, 534

\bibitem[{{M{\"u}ller} {et~al.}(2021){M{\"u}ller}, {Pawlowski}, {Lelli},
  {Fahrion}, {Rejkuba}, {Hilker}, {Kanehisa}, {Libeskind}, \&
  {Jerjen}}]{Muller2021b}
{M{\"u}ller}, O., {Pawlowski}, M.~S., {Lelli}, F., {et~al.} 2021, \aap, 645, L5

\bibitem[{{M{\"u}ller} {et~al.}(2018{\natexlab{b}}){M{\"u}ller}, {Rejkuba}, \&
  {Jerjen}}]{MuellerTRGB2018}
{M{\"u}ller}, O., {Rejkuba}, M., \& {Jerjen}, H. 2018{\natexlab{b}}, \aap, 615,
  A96

\bibitem[{{M{\"u}ller} {et~al.}(2019){M{\"u}ller}, {Rejkuba}, {Pawlowski},
  {Ibata}, {Lelli}, {Hilker}, \& {Jerjen}}]{Muller2019}
{M{\"u}ller}, O., {Rejkuba}, M., {Pawlowski}, M.~S., {et~al.} 2019, \aap, 629,
  A18

\bibitem[{{M{\"u}ller} {et~al.}(2017){M{\"u}ller}, {Scalera}, {Binggeli}, \&
  {Jerjen}}]{2017A&A...602A.119M}
{M{\"u}ller}, O., {Scalera}, R., {Binggeli}, B., \& {Jerjen}, H. 2017, \aap,
  602, A119

\bibitem[{{Oehm} {et~al.}(2017){Oehm}, {Thies}, \&
  {Kroupa}}]{2017MNRAS.467..273O}
{Oehm}, W., {Thies}, I., \& {Kroupa}, P. 2017, \mnras, 467, 273

\bibitem[{{Okamoto} {et~al.}(2015){Okamoto}, {Arimoto}, {Ferguson}, {Bernard},
  {Irwin}, {Yamada}, \& {Utsumi}}]{2015ApJ...809L...1O}
{Okamoto}, S., {Arimoto}, N., {Ferguson}, A. M.~N., {et~al.} 2015, \apjl, 809,
  L1

\bibitem[{{Okamoto} {et~al.}(2019){Okamoto}, {Arimoto}, {Ferguson}, {Irwin},
  {Bernard}, \& {Utsumi}}]{2019ApJ...884..128O}
{Okamoto}, S., {Arimoto}, N., {Ferguson}, A. M.~N., {et~al.} 2019, \apj, 884,
  128

\bibitem[{{Paudel} {et~al.}(2021){Paudel}, {Yoon}, \&
  {Smith}}]{2021ApJ...917L..18P}
{Paudel}, S., {Yoon}, S.-J., \& {Smith}, R. 2021, \apjl, 917, L18

\bibitem[{{Pawlowski}(2018)}]{Pawlowski2018}
{Pawlowski}, M.~S. 2018, Modern Physics Letters A, 33, 1830004

\bibitem[{{Pawlowski}(2021)}]{2021NatAs...5.1185P}
{Pawlowski}, M.~S. 2021, Nature Astronomy, 5, 1185

\bibitem[{{Pawlowski} {et~al.}(2017){Pawlowski}, {Ibata}, \&
  {Bullock}}]{2017ApJ...850..132P}
{Pawlowski}, M.~S., {Ibata}, R.~A., \& {Bullock}, J.~S. 2017, \apj, 850, 132

\bibitem[{{Pawlowski} \& {Kroupa}(2020)}]{Pawlowski2020}
{Pawlowski}, M.~S. \& {Kroupa}, P. 2020, \mnras, 491, 3042

\bibitem[{{Pawlowski} {et~al.}(2011){Pawlowski}, {Kroupa}, \& {de
  Boer}}]{2011A&A...532A.118P}
{Pawlowski}, M.~S., {Kroupa}, P., \& {de Boer}, K.~S. 2011, \aap, 532, A118

\bibitem[{{Pawlowski} {et~al.}(2012){Pawlowski}, {Pflamm-Altenburg}, \&
  {Kroupa}}]{2012MNRAS.423.1109P}
{Pawlowski}, M.~S., {Pflamm-Altenburg}, J., \& {Kroupa}, P. 2012, \mnras, 423,
  1109

\bibitem[{Pedregosa {et~al.}(2011)Pedregosa, Varoquaux, Gramfort, Michel,
  Thirion, Grisel, Blondel, Prettenhofer, Weiss, Dubourg, Vanderplas, Passos,
  Cournapeau, Brucher, Perrot, \& Duchesnay}]{scikit-learn}
Pedregosa, F., Varoquaux, G., Gramfort, A., {et~al.} 2011, Journal of Machine
  Learning Research, 12, 2825

\bibitem[{{Pillepich} {et~al.}(2018){Pillepich}, {Springel}, {Nelson}, {Genel},
  {Naiman}, {Pakmor}, {Hernquist}, {Torrey}, {Vogelsberger}, {Weinberger}, \&
  {Marinacci}}]{2018MNRAS.473.4077P}
{Pillepich}, A., {Springel}, V., {Nelson}, D., {et~al.} 2018, \mnras, 473, 4077

\bibitem[{{Planck Collaboration} {et~al.}(2016){Planck Collaboration}, {Ade},
  {Aghanim}, {Arnaud}, {Ashdown}, {Aumont}, {Baccigalupi}, {Banday},
  {Barreiro}, {Bartlett}, \& et~al.}]{2016A&A...594A..13P}
{Planck Collaboration}, {Ade}, P.~A.~R., {Aghanim}, N., {et~al.} 2016, \aap,
  594, A13

\bibitem[{{Poulain} {et~al.}(2021){Poulain}, {Marleau}, {Habas}, {Duc},
  {S{\'a}nchez-Janssen}, {Durrell}, {Paudel}, {Ahad}, {Chougule}, {M{\"u}ller},
  {Lim}, {B{\'\i}lek}, \& {Fensch}}]{2021MNRAS.506.5494P}
{Poulain}, M., {Marleau}, F.~R., {Habas}, R., {et~al.} 2021, \mnras, 506, 5494

\bibitem[{{Sales} {et~al.}(2022){Sales}, {Wetzel}, \&
  {Fattahi}}]{2022NatAs...6..897S}
{Sales}, L.~V., {Wetzel}, A., \& {Fattahi}, A. 2022, Nature Astronomy, 6, 897

\bibitem[{{Sawala} {et~al.}(2022){Sawala}, {Cautun}, {Frenk}, {Helly},
  {Jasche}, {Jenkins}, {Johansson}, {Lavaux}, {McAlpine}, \&
  {Schaller}}]{2022arXiv220502860S}
{Sawala}, T., {Cautun}, M., {Frenk}, C.~S., {et~al.} 2022, arXiv e-prints,
  arXiv:2205.02860

\bibitem[{{Sharina} {et~al.}(2001){Sharina}, {Karachentsev}, \&
  {Burenkov}}]{2001A&A...380..435S}
{Sharina}, M.~E., {Karachentsev}, I.~D., \& {Burenkov}, A.~N. 2001, \aap, 380,
  435

\bibitem[{{Spekkens} {et~al.}(2014){Spekkens}, {Urbancic}, {Mason}, {Willman},
  \& {Aguirre}}]{2014ApJ...795L...5S}
{Spekkens}, K., {Urbancic}, N., {Mason}, B.~S., {Willman}, B., \& {Aguirre},
  J.~E. 2014, \apjl, 795, L5

\bibitem[{Taibi {et~al.}(2023)Taibi, Pawlowski, Khoperskov, Steinmetz, \&
  Libeskind}]{taibi2023portrait}
Taibi, S., Pawlowski, M.~S., Khoperskov, S., Steinmetz, M., \& Libeskind, N.~I.
  2023, A portrait of the Vast Polar Structure as a young phenomenon: hints
  from its member satellites

\bibitem[{{Tikhonov} \& {Karachentsev}(1993)}]{1993A&A...275...39T}
{Tikhonov}, N.~A. \& {Karachentsev}, I.~D. 1993, \aap, 275, 39

\bibitem[{{Tully} {et~al.}(2013){Tully}, {Courtois}, {Dolphin}, {Fisher},
  {H{\'e}raudeau}, {Jacobs}, {Karachentsev}, {Makarov}, {Makarova},
  {Mitronova}, {Rizzi}, {Shaya}, {Sorce}, \& {Wu}}]{2013AJ....146...86T}
{Tully}, R.~B., {Courtois}, H.~M., {Dolphin}, A.~E., {et~al.} 2013, \aj, 146,
  86

\bibitem[{{Tully} {et~al.}(2015){Tully}, {Libeskind}, {Karachentsev},
  {Karachentseva}, {Rizzi}, \& {Shaya}}]{2015ApJ...802L..25T}
{Tully}, R.~B., {Libeskind}, N.~I., {Karachentsev}, I.~D., {et~al.} 2015,
  \apjl, 802, L25

\bibitem[{{{\v{Z}}emaitis} {et~al.}(2023){{\v{Z}}emaitis}, {Ferguson},
  {Okamoto}, {Cuillandre}, {Stone}, {Arimoto}, \&
  {Irwin}}]{2023MNRAS.518.2497Z}
{{\v{Z}}emaitis}, R., {Ferguson}, A. M.~N., {Okamoto}, S., {et~al.} 2023,
  \mnras, 518, 2497

\bibitem[{{Wang} {et~al.}(2020){Wang}, {Hammer}, {Rejkuba}, {Crnojevi{\'c}}, \&
  {Yang}}]{2020MNRAS.498.2766W}
{Wang}, J., {Hammer}, F., {Rejkuba}, M., {Crnojevi{\'c}}, D., \& {Yang}, Y.
  2020, \mnras, 498, 2766

\bibitem[{{Weisz} {et~al.}(2008){Weisz}, {Skillman}, {Cannon}, {Dolphin},
  {Kennicutt}, {Lee}, \& {Walter}}]{2008ApJ...689..160W}
{Weisz}, D.~R., {Skillman}, E.~D., {Cannon}, J.~M., {et~al.} 2008, \apj, 689,
  160

\bibitem[{{Yun} {et~al.}(1994){Yun}, {Ho}, \& {Lo}}]{1994Natur.372..530Y}
{Yun}, M.~S., {Ho}, P.~T.~P., \& {Lo}, K.~Y. 1994, \nat, 372, 530

\end{thebibliography}

\begin{table*}[ht]
\caption{The M81 satellite system.}             % title of Table
\centering                          % used for centering table
\begin{tabular}{l c c c r c l}        % centered columns (4 columns)
\hline\hline                 % inserts double horizontal lines
name &$\alpha_{2000}$ & $\delta_{2000}$ & $D$ & $M_r$ & $v$ & type  \\    % table heading 
 & & & Mpc & mag & km/s &   \\    % table heading 
\hline      \\[-2mm]                  % inserts single horizontal line
d0926+70 & 09:26:27.9 & +70:30:19 & 3.4$_{-0.2}^{+0.2}$ & -9.8 & --- & Tr   \\
d0934+70 & 09:34:03.22 & +70:12:58 & 3.0$_{-0.2}^{+1.1}$ & -9.4 & --- & dSph   \\
d0939+71 & 09:39:16.01 & +71:18:41 & 3.7$_{-0.4}^{+0.5}$ & -9.4 & --- & dSph   \\
HolmI & 09:40:32.3 & +71:11:11 & 4.0$_{-0.1}^{+0.1}$ & -14.6 & 139.4$\pm$0.1 & Ir   \\
d0944+69 & 09:44:22.49 & +69:12:38 & 3.8$_{-0.3}^{+0.6}$ & -6.8 & --- & dSph   \\
d0944+71 & 09:44:34.37 & +71:28:56 & 3.4$_{-0.1}^{+0.1}$ & -12.4 & --- & dSph   \\
F8D1 & 09:44:47.1 & +67:26:19 & 3.8$_{-0.1}^{+0.1}$ & -13.2 & -125.0$\pm$130.0 & dSph   \\
FM1 & 09:45:10.0 & +68:45:54 & 3.8$_{-0.1}^{+0.1}$ & -11.7 & --- & dSph   \\
NGC2976 & 09:47:15.6 & +67:54:49 & 3.7$_{-0.1}^{+0.1}$ & -18.0 & 6.0$\pm$4.0 & SAa   \\
KK77 & 09:50:10.0 & +67:30:24 & 3.8$_{-0.0}^{+0.0}$ & -13.0 & --- & dSph   \\
BK3N & 09:53:48.5 & +68:58:09 & 4.2$_{-0.2}^{+0.3}$ & -10.3 & -40.0$\pm$0.0 & Ir   \\
d0955+70 & 09:55:14.14 & +70:24:26 & 3.4$_{-0.4}^{+0.6}$ & -9.8 & --- & dSph   \\
KDG61 & 09:57:02.7 & +68:35:30 & 3.7$_{-0.0}^{+0.0}$ & -13.3 & 221.0$\pm$3.0 & dSph   \\
A0952+69 & 09:57:29.0 & +69:16:20 & 3.9$_{-0.3}^{+0.3}$ & nan & 99.0$\pm$0.0 & Ir   \\
HolmIX & 09:57:32.4 & +69:02:35 & 3.8$_{-0.1}^{+0.1}$ & -13.9 & 50.0$\pm$4.0 & Ir   \\
d0958+66 & 09:58:48.74 & +66:50:57 & 3.8$_{-0.1}^{+0.1}$ & -13.2 & 90.0$\pm$50.0 & BCD   \\
d0959+68 & 09:59:34.90 & +68:39:26 & 4.2$_{-0.3}^{+0.3}$ & -11.9 & --- & Tr   \\
NGC3077 & 10:03:21.0 & +68:44:02 & 3.8$_{-0.1}^{+0.1}$ & -17.8 & 19.0$\pm$4.0 & dSph   \\
Garland & 10:03:42.0 & +68:41:36 & 3.8$_{-0.4}^{+0.4}$ & nan & 43.0$\pm$17.0 & Ir   \\
BK5N & 10:04:40.3 & +68:15:20 & 3.7$_{-0.2}^{+0.2}$ & -11.7 & --- & dSph   \\
KDG63 & 10:05:07.3 & +66:33:18 & 3.6$_{-0.0}^{+0.0}$ & -13.0 & -129.0$\pm$0.3 & Tr   \\
d1006+67 & 10:06:46.80 & +67:12:00 & 3.6$_{-0.2}^{+0.2}$ & -9.8 & --- & dSph   \\
KDG64 & 10:07:01.9 & +67:49:39 & 3.8$_{-0.0}^{+0.0}$ & -13.2 & -15.0$\pm$13.0 & dSph   \\
IKN & 10:08:05.9 & +68:23:57 & 3.8$_{-0.0}^{+0.0}$ & -12.4 & -140.0$\pm$64.0 & dSph   \\
d1012+64 & 10:12:48.41 & +64:06:26 & 3.7$_{-0.1}^{+0.1}$ & -13.3 & 150.0$\pm$50.0 & BCD   \\
d1014+68 & 10:14:55.80 & +68:45:30 & 3.8$_{-0.3}^{+0.3}$ & -9.4 & --- & dSph   \\
d1015+69 & 10:15:06.89 & +69:02:14 & 3.9$_{-0.2}^{+0.3}$ & -8.8 & --- & dSph   \\
HS117 & 10:21:25.2 & +71:06:58 & 4.0$_{-0.1}^{+0.1}$ & -12.1 & -37.0$\pm$0.0 & Tr   \\
DDO78 & 10:26:27.9 & +67:39:24 & 3.5$_{-0.0}^{+0.0}$ & -12.8 & 55.0$\pm$10.0 & dSph   \\
IC2574 & 10:28:22.4 & +68:24:58 & 3.9$_{-0.0}^{+0.0}$ & -17.7 & 43.0$\pm$4.0 & T8   \\
d1028+70 & 10:28:39.98 & +70:14:00 & 3.8$_{-0.1}^{+0.1}$ & -12.4 & -114.0$\pm$50.0 & BCD   \\
DDO82 & 10:30:35.0 & +70:37:10 & 3.9$_{-0.0}^{+0.0}$ & -15.1 & 56.0$\pm$3.0 & Im   \\
BK6N & 10:34:31.9 & +66:00:42 & 3.3$_{-0.2}^{+0.2}$ & -11.7 & --- & dSph   \\
d1041+70 & 10:41:18.14 & +70:09:14 & 3.7$_{-0.3}^{+0.2}$ & -9.3 & --- & dSph   \\
\hline
\end{tabular}
\tablefoot{ The references for the original distances are: Holm I \citep{2002A&A...383..125K}, F8D1 \citep{2000A&A...363..117K}, FM1 \citep{2001A&A...375..359K}, NGC2976 \citep{2002A&A...383..125K}, KK77 \citep{2000A&A...363..117K}, BK3N \citep{2002A&A...383..125K}, KDG061 \citep{2000A&A...363..117K}, A0952+69 \citep{2002A&A...383..125K}, Holm IX \citep{2009ApJS..183...67D}, NGC3077 \citep{2003A&A...398..479K}, Garland \citep{2002A&A...383..125K}, BK5N \citep{2000A&A...363..117K}, KDG063 \citep{2000A&A...363..117K}, KDG064 \citep{2000A&A...363..117K}, IKN \citep{2006AJ....131.1361K}, HS117 \citep{2006AJ....131.1361K}, DDO078 \citep{2000A&A...363..117K}, IC2574 \citep{2002A&A...383..125K}, DDO82 \citep{2002A&A...383..125K}, and BK6N \citep{2000A&A...363..117K}.
The references for the velocities are: Holm I \citep{2003A&A...401..483H}, F8D1 \citep{2009AJ....137.3009C}, NGC2976 \citep{1981MNRAS.195..327A}, BK3N \citep{1993A&A...275...39T}, KDG061 \citep{2010MNRAS.406.1152M}, A0952+69 \citep{2003MNRAS.342..738L}, Holm IX \citep{1981MNRAS.195..327A}, d0958+66 \citep{2009AJ....137.3009C}, NGC3077 \citep{1981MNRAS.195..327A}, Garland \citep{1985MNRAS.217..731K}, KDG063 \citep{2003A&A...401..483H}, KDG064 \citep{2010MNRAS.406.1152M}, IKN \citep{2009AJ....137.3009C}, d1012+64 \citep{2009AJ....137.3009C}, HS117 \citep{2004AJ....127.2031K}, DDO078 \citep{2001A&A...380..435S}, IC2574 \citep{1981MNRAS.195..327A}, d1028+70 \citep{2009AJ....137.3009C}, and DDO82 \citep{2003A&A...405..951M}. Note that for A0952+69 and HS117 no uncertainty is given for the velocity estimate. The morphological type is according to the Local Volume catalog \citep{2013AJ....145..101K}.}
\label{tab:sample}
\end{table*}

\end{document}